\documentclass[12pt]{article}
\usepackage{epsfig}

\usepackage{a4}
\usepackage{amsmath}

\usepackage{amssymb}
\usepackage{graphicx}
\usepackage{amsmath}
\usepackage{amsfonts} 

\newcommand{\la}{\lambda} 
\newcommand{\La}{\Lambda}
\newcommand{\ka}{\kappa}
\newcommand{\f}{\phi}
\newcommand{\F}{\Phi}
\newcommand{\vf}{\varphi}
\newcommand{\ta}{\theta}
\newcommand{\Ta}{\Theta}
\newcommand{\al}{\alpha}
\newcommand{\bt}{\beta}
\newcommand{\ga}{\gamma}
\newcommand{\de}{\delta}
\newcommand{\si}{\sigma}
\newcommand{\Si}{\Sigma}

\newcommand{\ee}{\end{equation}}
\newcommand{\eea}{\end{eqnarray}}
\newcommand{\be}{\begin{equation}}
\newcommand{\bea}{\begin{eqnarray}}

\newcommand{\pa}{\partial}
\newcommand{\Om}{\Omega}
\newcommand{\vep}{\varepsilon}
\newcommand{\vr}{\varrho}

\newcommand{\om}{\omega}
\def\theequation{\arabic{equation}}

\newcommand{\re}[1]{(\ref{#1})}
\newcommand{\R}{{\rm I \hspace{-0.52ex} R}}

\newcommand{\eins}{1\hspace{-0.56ex}{\rm I}}

\def\theequation{\thesection.\arabic{equation}}

\begin{document}

\title{\bf The role of the Callan-Witten anomaly density 
as a Chern-Simons term in Skyrme model}

\author{ {\large Francisco Navarro-L\'erida}$^{\star}$, 
{\large Eugen Radu}$^{\diamond}$ 
and 
{\large D. H. Tchrakian}$^{\dagger **}$
\\ 
\\
$^{\star}${\small  
Departamento de F\'isica Te\'orica and IPARCOS, Ciencias F\'isicas,
}\\
{\small Universidad Complutense de Madrid, E-28040 Madrid, Spain} 
\\  
$^{\diamond}${\small
 Departamento de Matem\'atica da Universidade de Aveiro and }
\\
{\small 
Center for Research and Development in Mathematics and Applications,}
\\
{\small 
 Campus de Santiago, 3810-183 Aveiro, Portugal}
\\ 
$^{\dagger}${\small 
School of Theoretical Physics, Dublin Institute for Advanced Studies,}
\\
{\small  Burlington Road, Dublin 4, Ireland}
\\   
$^{**}${\small Department of Computer Science, National University of Ireland Maynooth, Maynooth, Ireland }
}

\maketitle

\date{}

 \begin{abstract}

We consider axially symmetric solutions of the U(1) gauged Skyrme model
supplemented with a  Callan-Witten (CW) anomaly density term.
The main properties of the solutions are studied,  several specific features 
introduced by the presence of the CW term being identified.
We find that 
 the solitons  possess a nonzero angular momentum proportional to the electric charge,
which in addition to the usual Coulomb part,
acquires an extra (topological) contribution from the CW term.
Specifically, it is shown that the slope of mass/energy $M$ $vs.$ electric charge $Q_e$
and angular momentum $J$ can be both positive $and$ negative.
% a feature that is found
%in cases where the Chern-Simons (CS) term is used in corresponding odd
%dimensional spacetimes. 
Furthermore, it is shown that the gauged Skyrmion
persists even when the quartic (Skyrme) kinetic term disappears.
\end{abstract}

\bigskip
\bigskip
\bigskip
\bigskip
\bigskip
\bigskip
\bigskip

\noindent
{\it We dedicate this work to the memory of our colleague Valery Rubakov with whom we discussed this topic extensively}

%\ \ \ PACS Numbers: 04.50.+h, 11.10.Kk, 11.15.Kc

%\tableofcontents

\newpage
%%%%%%%%%%%%%%%%%%%%%%%%%%%%%%%%%%%%%%%%%%
\section{Introduction}
%%%%%%%%%%%%%%%%%%%%%%%%%%%%%%%%%%%%%%%%%%
Solutions of the classical Skyrme model \cite{Skyrme:1961vq,Skyrme:1961vr} were proposed as the solitons describing the
dynamics of nucleons a long time ago. Later, topological aspects of the model were highligted in
Ref.~\cite{Balachandran:1988sn,Balachandran:1982dw},
followed by the influential work of Ref.~\cite{Witten:1983tw}. Subsequently, a detailed study of the $U(1)$ gauged Skyrme model
was presented in Ref.~\cite{Callan:1983nx}.

More recently, $SO(2)$ gauged Skyrmions~\footnote{
The term Skyrmion here is used for the soliton of a $O(D+1)$ sigma model on $\R^D$ in
$D+1$ dimensional space-time.} in $3+1$ dimensions
were studied quantitatively in Refs.~\cite{Piette:1997ny,Radu:2005jp}
(see also \cite{Livramento:2023keg}).
These solitons carry both electric charge $Q_e$~\cite{Piette:1997ny}  and angular
 momentum $J$~\cite{Radu:2005jp}, and are characterised by the
asymptotic value of the electric component $A_0$ of the Maxwell potential $A_\mu$. It was observed
that the the mass/energy $M$ of the Abelian gauged Skyrmion increases monotonically with
$Q_e$ and $J$.

Subsequently the effect of the Chern-Simons density on gauged Skyrmions on Abelian gauged Skyrme scalars in $2+1$ 
dimensions~\cite{Navarro-Lerida:2016omj,Navarro-Lerida:2018siw}, and in $4+1$ 
dimensions~\cite{Navarro-Lerida:2020jft,Navarro-Lerida:2020hph} was studied. In these
particular examples, the solutions are characterised by the
asymptotic value of the magnetic component $A_i$ as well as that of $A_0$ as seen in~\cite{Piette:1997ny,Radu:2005jp}. It turns out that in the presence of the Chern-Simons
density in the Lagrangian, the $(M,Q_e)$ and $(M,J)$ curves exhibit both positive and negative
slopes, in contrast to the monotonic positive slopes observed
in~\cite{Piette:1997ny,Radu:2005jp} in the absence of a Chern-Simon term.

Our aim in the present work is to test whether the $U(1)$ gauged Skyrmions in $3+1$ dimensions
exhibit the new pattern of negative (as well as positive) slopes of $(M,Q_e)$ and $(M,J)$ curves when the Lagrangian
is augmented with a Chern-Simons--like term, namely the Anomaly related term
employed in Ref.~\cite{Callan:1983nx}, whose construction is given in  Ref.~\cite{Witten:1983tw} on global anomalies.
Our results show that the answer to this question is affirmative!

In Section {\bf 2} we introduce the model, in Section {\bf 3} we define the global charges in
this model, in Section {\bf 4} the system is subjected to axial symmetry and in Section {\bf 5}
the numerical construction is presented in detail. We have found it convenient to use a
constraint compliant parametrisation of the Skyrme scalar in Sections {\bf 3} and {\bf 4},
while in {\bf 5} the Lagrange--multiplier method is used in the general parametrisation. The
constraint-compliant parametrisation employed here is presented in Appendix {\bf A}, and 
in Appendix {\bf B} an alternative such parametrisation is given.

\medskip
%%%%%%%%%%%%%%%%%%%%%%%%%%%%%%%%%%%%%%%%%%%%%%%%%%%%%%%%%%%%%%%%%%%%%%%%% 
 {\bf Conventions.}
%%%%%%%%%%%%%%%%%%%%%%%%%%%%%%%%%%%%%%%%%%%%%%%%%%%%%%%%%%%%%%%%%%%%%%%%%
%
Throughout the paper, Greek alphabet letters $\alpha,\beta,\dots$ 
label spacetime coordinates, 
  running from $0$ to $3$ (with $x^0=t$);
early Latin letters, $a,b,\dots $ label the internal indices of the scalar field multiplet.
As standard, we use Einstein's summation convention, but to alleviate  notation, no distinction is made between covariant and contravariant \textit{internal} indices.

The background of the theory is Minkowski spacetime (with the signature $(+---)$), 
where the spatial $\mathbb{R}^3$ is written
first 
 in terms of 
Cartesian coordinate, 
$ds_3^2=dx_1^2+dx_2^2+dx_3^2$,
(with $x_3=z$). 
The transformation
$x_1=\rho \cos \varphi$
$x_2=\rho \sin \varphi$
leads to metric of flat space in
cylindrical coordinates,
%\begin{eqnarray}
%\label{metric}
%
$ds_3^2=d\rho^2+\rho^2 d\varphi^2+dz^2,$
%\end{eqnarray}
where
$0\leq \rho<\infty$,
$-\infty<z<\infty$
and 
$0\leq \varphi <2\pi$.
Finally,
we shall use also
the flat space metric in spherical coordinates,
$
ds_3^2=dr^2+r^2(d\theta^2+\sin^2\theta d\varphi^2),
$
obtained via
the coordinate transformation
$\rho=r\sin \theta$,
$z=r\cos \theta$
(with $0\leq r <\infty$
and $0\leq \theta \leq \pi$).

%%%%%%%%%%%%%%%%%%%%%%%%%%%%%%%%%%%%%%%%%%%%%%
\section{The model}
%%%%%%%%%%%%%%%%%%%%%%%%%%%%%%%%%%%%%%%%%%%%%

The model we study is that considered in Ref.~\cite{Callan:1983nx} described by the Lagrangian
\bea
\label{Lin}
L&=&\frac14\,F_{\mu\nu}F^{\mu\nu}+L_{\rm Skyrme}
+ \kappa~ \Om_{\rm CW},\label{U1Sk1}
\\
L_{\rm Skyrme}&=&-\frac12\,\mbox{Tr}\,D_\mu U\,D^\mu U^{-1}+\frac14\,\mbox{Tr}[U^{-1}D_\mu U,U^{-1}D_\nu U]^2\label{U1Sk2}
+V[U,U^{-1}]
\eea
in which the Skyrme scalar is parametrised by the $SU(2)$ group element $U$ and $V[U,U^{-1}]$ is a potential term
~\footnote
{  This potential term is not essential since the Derrick scaling requirement is satisfied in its absence. It is
included for convenience in the numerical construction by supplying exponential decay.}.

It is convenient to express the $\Om_{\rm CW}$ term as
\bea
\Om_{\rm CW}&=&\Om_{\rm CW}^{(1)}+\Om_{\rm CW}^{(2)}\label{anomaly12}\\
&\equiv&A_\tau\,W_{(1)}^{\tau}+A_\tau F_{\mu\nu}\,W_{(2)}^{\tau\mu\nu}\label{anomaly1212}
\eea
where $F_{\mu\nu}=\pa_\mu A_\nu-\pa_\nu A_\mu$ and
\bea
W_{(1)}^{\tau}&=&\vep^{\tau\la\mu\nu}\,\mbox{Tr}Q\left[(U^{-1}\pa_\mu U)(U^{-1}\pa_\nu U)(U^{-1}\pa_\la U)+
(U\pa_\mu U^{-1})(U\pa_\nu U^{-1})(U\pa_\la U^{-1})\right] ,
\label{W1}
\\
W_{(2)}^{\tau\mu\nu}&=&i\,\vep^{\tau\la\mu\nu}\left\{\mbox{Tr}\left[Q^2(U^{-1}\pa_\la U+U\pa_\la U^{-1})\right]
+\frac12\,\mbox{Tr}\left[(Q\pa_\la UQU^{-1}-Q\pa_\la U^{-1}QU)\right]\right\}.
\label{W2}
\eea

The covariant derivative in \re{U1Sk1} is defined by\footnote{Note that 
the definition in
\cite{Callan:1983nx}
is
$
D_\mu U=\pa_\mu U+i e A_\mu[Q,U]
$
with $e$
the gauge coupling constant,
which, in order to simplify various relations, we set to one.
Also, this is the value considered in Section 5 dealing with the numerical results.
}
\be
\label{covU}
D_\mu U=\pa_\mu U+iA_\mu[Q,U]\ ,\quad {\rm with}\quad Q=\frac13\left(
\begin{array}{cc}
2\ & 0\\
0\ & -1
\end{array}
\right)\,.
\ee

The alternative to the formulation of the Skyrme model in terms of the $SU(2)$ element $U$ is the
formulation in terms of the Skyrme scalar $\f^a=(\f^I,\f^4),\ I=1,2,3$ of the $O(4)$ sigma model satisfying the
constraint
\[
|\f^a|^2=1\ \quad \Leftrightarrow \ \quad U^{\dagger}U=\eins\,,
\]
expressed through the relation(s)
\be
\label{fU}
U=\f^4\,\eins+i\,\f^I\si_I\ ,\quad U^{-1}=\f^4\,\eins-i\,\f^I\si_I\ ,\quad I=1,2,3\,,
\ee
with the $O(4)$ Skyrme scalar $\f^a=(\f^I,\f^4)$.

Reassigning the index $a$ as $a=(\al,A)$, with $\al=1,2$ and $A=3,4$,  
 the following useful relations can be stated
\be
[Q,U]=-\vep_{\al\bt}\,\f^\al\,\si^\bt\,,\label{QU}\
\ee
\be
[Q,U^{-1}]=\vep_{\al\bt}\,\f^\al\,\si^\bt\,,\label{QU-}
\ee
of which \re{QU} then leads to the $SO(2)$ gauging prescription
\bea
D_\mu\f^{\al}&=&\pa_\mu\f^{\al}+A_\mu(\vep\f)^{\al}\ ,\quad\al=1,2\label{coval}\\
D_\mu\f^{A}&=&\pa_\mu\f^A\label{covA}\ ,\qquad\qquad\quad A=3,4\,,
\eea
for the $O(4)$ Skyrme scalar $\f^a=(\f^\al,\f^A)$ employed in \cite{Tchrakian:2015pka}~\footnote{In Ref.~\cite{Callan:1983nx}, 
the definition used for the covariant derivative of $U$ is
\be
\label{covUCW}
D_\mu U=\pa_\mu U-iA_\mu[Q,U]
\ee
in which case the definition of the covariant derivative $D_\mu\f^a$ given by \re{coval}-\re{covA}, used in
\cite{Tchrakian:2015pka}, changes to
\bea
D_i\f^{\al}&=&\pa_i\f^{\al}-A_i(\vep\f)^{\al}\ ,\quad\al=1,2\label{coval2CW}
\\
D_i\f^{A}&=&\pa_i\f^A\label{covaA2CW}\ ,\qquad\qquad\quad A=3,4\,.
\eea
}
which through \re{fU}, is precisely equivalent to the definition \re{covU}.

Here, we will choose to use the definition \re{covU} for $D_\mu U$, consistently with the definition $D_\mu\f^a$
given by \re{coval}-\re{covA}. Further abbreviating
our notation for $D_\mu\f^a=(D_\mu\f^\al,D_\mu\f^A)$
\[
\f_\mu^a\stackrel{\rm def.}=D_\mu\f^a\ ,\quad{\rm and}\quad\f_{\mu\nu}^{ab}\stackrel{\rm def.}=
\f_{[\mu}^a\f_{\nu]}^b
\]
the Skyrme Lagrangian  
\re{U1Sk2} can be written compactly as
\be
\label{Skyrme}
L_{\rm Skyrme}=-\frac12\,|\f_\mu^a|^2+\frac18\,|\f_{\mu\nu}^{ab}|^2+V[\f^a]
\ee
where $V[\f^a]$ is the (pion mass) potential term intended to localise the soliton exponentially.

With this parametrisation of the Skyrme scalar, the stress-energy tensor $T_{\mu\nu}$ can be expressed compactly
 as
\bea
T_{\tau\la}&=&-\frac12\left(F_{\tau\mu}F_\la{}^\mu-\frac14\,g_{\tau\la}\,|F_{\mu\nu}|^2\right)
+\frac12\,
%\eta^2
\left(\f_\tau^a\,\f_\la^a-\frac12\,g_{\tau\la}\,|\f_\mu^a|^2\right)\nonumber\\
&&\qquad\qquad\qquad-\frac14\left(\f_{\tau\mu}^{ab}(\f_\la{}^\mu)^{ab}-\frac14\,g_{\tau\la}\,|\f_{\mu\nu}^{ab}|^2\right)\label{stress}\,.
\eea

The component $T_{00}$ is the energy density, 
while the angular momentum density $\cal J$ in the $(x_1,x_2)$-plane is
\bea
\label{calJ}
{\cal J}&\stackrel{\rm def.}=&x_2\,T_{10}-x_1\,T_{20}\nonumber\\
&=&(\vep x)_{\bar\al}\,T_{\bar\al 0}~,
\eea
in which the spacelike index notation ($\bar\al=1,2$) is used. The quantity $T_{\bar\al 0}$ is
\be
T_{\bar\al 0}=\frac12(F_{\bar\al\bar\bt}F_{0\bar\bt}+F_{\bar\al z}F_{0z})+\frac12\,\f_{\bar\al}^a\f_0^a+
\frac12(\f_{\bar\al\bar\bt}^{ab}\f_{0\bar\bt}^{ab}+\f_{\bar\al z}^{ab}\f_{0z}^{ab})\label{stress1}
\ee
which can be evaluated only after imposition of axial symmetry.

Having introduced the re-expression of $U$ in terms of the Skyrme scalar $\f^a=(\f^I,\f^4)=(\f^\al,\f^A)$ with $I=1,2,3$ and
($\al=1,2;\,A=3,4)$, one exploits the relation \re{QU-}, and one finds a useful simplification of $W_{(2)}^{\tau\mu\nu}$ in \re{W2}
by noting
\bea
\mbox{Tr}(
Q\pa_\la UQU^{-1}-Q\pa_\la U^{-1}QU)&=&\mbox{Tr}\,Q^2(U^{-1}\pa_\la U+\pa_\la UU^{-1})
+\vep_{\al\bt}\f^\al\mbox{Tr}\,\si^\bt Q\pa_\mu(U+U^{-1})\nonumber\\
&=&\mbox{Tr}\,Q^2(U^{-1}\pa_\la U+\pa_\la UU^{-1})
+\vep_{\al\bt}\,\f^\al\,\pa_\la\f^4\,\mbox{Tr}\,Q\,\si^\bt
\nonumber\\
&=&\mbox{Tr}\,Q^2(U^{-1}\pa_\la U+\pa_\la UU^{-1})\label{term}
\eea
since $\mbox{Tr}\,Q\,\si^\bt=0$, and hence the second term in \re{W2}
equals  $\frac12$ times the fist term there, $i.e.,$
\bea
W_{(2)}^{\tau\mu\nu}&=&
\frac{3i}{2}\,\vep^{\tau\la\mu\nu}\,\mbox{Tr}\left[Q^2(U^{-1}\pa_\la U
+\pa_\la UU^{-1})\right]\,,\label{W2n}
\eea
which simplifies the expression \re{W2}. This is of practical importance when expressing the Lagrangian using a
parametrisation in which the Skyrme scalar is compliant with the constraint $U^{\dagger}U=\eins$ (or $|\f^a|^2=1$).
Working with a constraint compliant parametrisation is very convenient in the definition of the electric charge, and, in
verifying that the Euler-Lagrange equations of $\Om_{\rm CW}$ are gauge invariant.%~\footnote{This serves in practice to ensure that the rather complicated expression of $\Om_{\rm CW}$ used is correct.}.

One such parametrisation, designated ``$3+1$'', is the one employed in the present work and is given in Appendix {\bf A}. An
alternative one, designated ``$2+2$'', is relegated to Appendix {\bf B}. Our preference for the particular
parametrisation of Appendix {\bf A} is dictated by technicalities of numerical constructions.

%%%%%%%%%%%%%%%%%%%%%%%%%%%%%%%%%%%%%%%%%%%%%%%%%%%%%%%%
\section{Definitions of global charge densities}
%%%%%%%%%%%%%%%%%%%%%%%%%%%%%%%%%%%%%%%%%%%%%%%%%%%%%%%%
In this Section, the definitions of the densities of global charges (baryon number and
electric charge) are presented, in preparation to applying axial
symmetry and their evaluatuation  in the next Section. Here, the definitions of the gauge deformed baryon number  
density of the electric charge are given explicitly, while the definition of the angular momentum
\re{calJ} given above in terms of components of the stress-energy tensor \re{stress}, will be made precise in the following Section
after imposition of symmetry.

\subsection{Gauge deformed baryon number: energy lower bound}
The baryon number $q$ of the nucleon is defined to be the winding number of the Skyrme scalar $\f^a=(\f^I,\f^4)$,
$I=1,2,3$, seen in the previous section. This is the volume integral of the winding number density
\bea
\vr_0&=&\frac13\,\vep_{ijk}\vep^{abcd}\,\pa_i\f^a\pa_j\f^b\pa_k\f^c\f^d\label{vr0x}\\
&=&\vep_{ijk}\vep^{IJK}\left[\pa_i\f^I \pa_j\f^J\pa_k\f^K\f^4-3\,\pa_i\f^I \pa_j\f^J\pa_k\f^4\f^K
\right]\label{vr10x}
\eea
and with the axially symmetric parametrisation
\be
\f^I=\left(\begin{array}{l}\sin f\ n^\al\\ \cos f\end{array}\right)
\quad{\rm with}\quad
n^\al=\left(\begin{array}{l}
\cos n\vf\\
\sin n\vf
\end{array}\right)\label{ax}
\ee
and boundary values $f(0)=\pi\,,f(\infty)=0$, the integral of $\vr_0$ equals the integer (winding number) $n$.
Most importantly, this number $q=n$ presents a lower bound~\footnote{
Though for $n > 1$ the actual energy of the Skyrmion is known from the work of \cite{Battye:1997qq},
to be somewhat less than that of the axially symmetric configuration \re{ax}.}
on the energy of the Skyrmion on $\R^3$.

After $SO(2)$ gauging according to \re{coval}-\re{covA}, the winding number density \re{vr0x} no longer
results in a lower bound on the energy of the gauged Skyrmion.

 It was proposed in Ref.~\cite{Tchrakian:2015pka} (and references therein) that the energy lower bound (density) replacing
the winding number density for a $O(D+1)$ Skyrme scalar on $\R^D$ suffers a deformation caused by the gauging,
 which must now be both total divergence like the latter, and gauge invariant.
 
These gauge deformed winding number densities are given by two equivalent definitions, formally expressed as
\bea
\vr&=&\vr_G+W[F,D\F]\label{vr11x}\\
&=&\vr_0+\pa_i\Om_i[A,\F]\label{vr21x}\,.
\eea

The definition \re{vr21x}, which consists of the explicitly total divergence term $\pa_i\Om_i[A,\F]$ and the
winding number density $\vr_0$ which is essentially total divergence, is total divergence. It is also gauge
invariant since it is equal to the gauge invariant version \re{vr11x}, in which $W[F,D\F]$ is by construction
gauge invariant, while $\vr_G$ is gauge invariant by virtue of its definition, which is that of $\vr_0$ with all
partial derivatives replaced by covariant derivatives. Thus, $\vr$ is both gauge invariant and total divergence.

%where $\vr_0$ in \re{vr21x} is the winding number density in the given dimensions and $\Om_i[A,\F]$ is gauge variant,
%and in \re{vr11x} $W[F,D\F]$ and $\vr_G$ are both explicitly gauge invariant, the definition of $\vr_G$ being that of
%$\vr_0$ with all partial derivatives replaced by covariant derivatives. Thus, the equivalent definitions \re{vr11x}
%and \re{vr21x} are both gauge invariant and total divergence as required of an energy lower bound.

%{\bf Eugen: please check this sentence.}

%one of which is explicitly gauge invariant and the other that is explicitly total divergence, both being gauge 
%invariant {\bf and} total divergence.

%The definition \re{vr11x} is explicitly gauge invariant and is suitable for establishing Bogomol'nyi like
%energy lower bounds, while \re{vr21x} is useful for tracking the deviation from the baryon number under the 
%influence of the gauge field, namely its departure from the winding number density $\vr_0$.

In the case of the $SO(2)$ gauged $O(4)$ sigma (Skyrme) model on $\R^3$ at hand, these two equivalent definitions
of $\vr$ are
\bea
\vr&=&\vr_G+\frac12\,\vep_{ijk}\,F_{ij}(\f^B \pa_k\f^A)\label{vr11}\\
&=&\vr_0+\vep_{ijk}\,\pa_i(A_j\,\vep^{AB}\f^B \pa_k\f^A)\label{vr21}
\eea
in which $\f^\al$, with $\al=1,2$ are the components of the $O(4)$ Skyrme scalar that are gauged with $SO(2)$ and
$\f^A$, with $A=3,4$ are the components that are not gauged. 

The quantity $\vr_0$ \re{vr0x}-\re{vr10x} can alternatively be expressed as
\bea
\vr_0%&=&\frac13\,\vep_{ijk}\vep^{abcd}\,\pa_i\f^a\pa_j\f^b\pa_k\f^c\f^d\label{vr0}\\
&=&\vep_{ijk}\left[(\vep^{\al\bt}\pa_i\f^\al \pa_j\f^\bt)(\vep^{AB}\f^B\,\pa_k\f^A)+
(\vep^{AB}\pa_i\f^A \pa_j\f^B)(\vep^{\al\bt}\f^\bt\,\pa_k\f^\al)\right]\label{vr10}
\eea
and the density $\vr_G$ is defined by \re{vr10}, with each partial derivative $\pa_i\f^\al$ replaced by the
covariant derivative $D_i\f^\al$.

%%%%%%%%%%%%%%%%%%%%%%%%%%%%%%%%%%%%%%%%%%
\subsection{Definition of the (electric) Noether charge}
%%%%%%%%%%%%%%%%%%%%%%%%%%%%%%%%%%%%%%%%%%

Consider the Lagrangian density
\bea
{\cal L}&=&\frac14\,|F_{\mu\nu}|^2-\frac12\,
%\eta^2\,
|D_\mu\f^a|^2
+\frac18\,|D_{[\mu}\f^aD_{\nu]}\f^b|^2+V[\f^4]+\ka\,\Om_{\rm CW},
\label{L24}
\\
\Om_{\rm CW}&=&\vep^{\tau\la\mu\nu}\,F_{\mu\nu}\bigg[-(g-\sin g)\pa_\tau\cos\Ta\,\pa_\la\F
+A_\tau(\pa_\la g\,\cos\Ta+\sin g\,\pa_\la\cos\Ta)\bigg]\label{LCW}
\eea
in which the anomaly term \re{LCW} is expressed explicitly in terms of the constraint compliant parametrisation \re{U} and \re{param} presented in
Appendix {\bf A}. 
The quadratic and quartic Skyrme kinetic terms in this parametrisation are given by
\re{quadSk} and \re{quartSk} respectively.
Also,
\bea
&&\de_{A_\tau}{\cal L}=-\pa_\mu\,F^{\mu\tau}-2\ka
\vep^{\tau\la\mu\nu}\bigg[(1-\cos g)\pa_\la g\,\pa_\mu\cos\Ta(A_\nu-\pa_\nu\F)\nonumber\\
&&\quad+F_{\mu\nu}(\pa_\la g\,\cos\Ta+\sin g\,\pa_\la\cos\Ta)\bigg]
-\frac12
%\eta^2
(1-\cos g)\sin^2\Ta\,(A^\tau-\pa^\tau\F)=0
\label{dA1}
\eea
and the equation resulting from the variation w.r.t. $\F$ is
\be
\de_{\F}{\cal L}=-\ka\,\vep^{\tau\la\mu\nu}
\,(1-\cos g)\pa_\tau g\ \pa_\la\cos\Ta\ F_{\mu\nu}-\frac12
%\eta^2\,
\pa_\tau\left[(1-\cos g)\sin^2\Ta\,(A^\tau-\pa^\tau\F)
\right]=0\,.
\label{dF1}
\ee
%and the equation resulting from the variation w.r.t. $\F$ is
%\be
%\de_{\F}{\cal L}=-\ka\,\vep^{\tau\la\mu\nu}
%\,(1-\cos g)\pa_\tau g\ \pa_\la\cos\Ta\ F_{\mu\nu}-\frac12\eta^2\,\pa_\tau\left[(1-\cos g)\sin^2\Ta\,(A^\tau-\pa^\tau\F)
%\right]=0\,.
%\label{dF1}
%\ee
(Note that in \re{dA1} and \re{dF1}, the terms coming from the variations of the {\it quartic} kinetic
term are omitted for typographic economy.)

We note that the divergence of the Maxwell equation \re{dA1} turns out to be precisely the $\F$
 equation \re{dF1}, so that the constraint imposed on the system by taking the divergence of the
Maxwell equation is satisfied by the $\F$ equation \re{dF1}.

The main difference between the gauged Skyrme models endowed with an Anomaly-type term as in
\re{LCW} and those with Chern-Simons (CS) terms is, that in the CS the term resulting from
the variation w.r.t. the gauge field is {\it divergenceless} while,
the corresponding term in \re{dA1} is {\bf not} divergenceless.

Expressing the Maxwell equation \re{dA1} as
\bea
&&-\pa_\mu\,F^{\mu\tau}-\ka\,\Om^\tau
=\frac12
%\eta^2
(1-\cos g)\sin^2\Ta\,(A^\tau-\pa^\tau\F)\label{dA11}\\
&&\Om^\tau=2\,\vep^{\tau\la\mu\nu}\left[F_{\mu\nu}(\pa_\la g\,\cos\Ta+\sin g\,\pa_\la\cos\Ta)+
(1-\cos g)\pa_\la g\,\pa_\mu\cos\Ta(A_\nu-\pa_\nu\F)\right]\label{Om}
\eea
it is clear that the current on the right hand side of \re{dA11} is not a conserved current since the divergnce of
$\Om^\ta$ does not vanish. Therefore the the $0$-th component of that current cannot define the electric charge.
Instead, the electric current must be defined using the Noether Theorem.

The symmetry of this system is the invariance under an Abelian gauge transformation
\bea
A_\mu\to\tilde A_\mu&=&A_\mu+\pa_\mu\La\label{Abel}\\
\F\to\tilde\F&=&\F+\La\label{La}
\eea
$\La$ being an arbitrary function.

One can now invoke the (first) Noether Theorem by considering a global transformation with $\La={\rm const.}$. That
gives a one parameter continuous symmmetry of the action. Since
\[
\frac{\de\F}{\de\La}=1\,,
\]
the divergenceless Noether current is defind by
\be
\label{Norther}
j_{(N)}^\mu=\frac{\pa\cal L}{\pa(\pa_\mu\F)}
\ee
which can be readily calculated for the Lagrangian \re{LCW} to give
\be
\label{Noetj}
j_{(N)}^\tau=-\frac12
%\eta^2
(1-\cos g)\sin^2\Ta(A^\tau-\pa^\tau\F)
+\ka(g-\sin g)\,\vep^{\tau\la\mu\nu}\,F_{\mu\nu}\,\pa_\la\cos\Ta\,.
\ee
%Substituting the term multiplying $\eta^2$ in the Maxwell equation \re{dA1}, 
By usinng the equation \re{dA1}, 
% \re{Noetj},
 one finds after
some manipulations
\be
\label{Noetjfin}
j_{(N)}^\tau=\pa_\mu F^{\mu\tau}+2\ka\,\vep^{\tau\la\mu\nu}\,\pa_\la\left[
(g-\sin g)\,\pa_\mu\cos\Ta\,(A_\nu-\pa_\nu\F)+F_{\mu\nu}\,g\,\cos\Ta\right]
\ee
which is manifestly a conserved 
current.

Equation \re{Noetjfin} plays the role of the Maxwell equation, and its $0$-th component in the static limit, which plays
the role of the Gauss Law equation 
\be
\label{NoetGauss}
\pa_i F_{i0}-2\ka\,\vep_{ijk}\,\pa_k\left[
(g-\sin g)\,\pa_i\cos\Ta\,(A_j-\pa_j\F)+F_{ij}\,g\,\cos\Ta\right]=j_{(N)}^0
\ee
in which $j_{(N)}^0$ plays the role of the zeroth component of the usual Maxwell current and the term with strength $\ka$
on the LHS plays the role of $\vep_{ij\dots mn}F_{ij}\dots F_{mn}$ occurring in the usual case for Chern-Simons in all odd
dimensions.

The electric charge is then defined as
\bea
Q_e&=&
%\frac{1}{4\pi}
\int\,j_{(N)}^0\,d^3x\stackrel{\rm def.}=Q_{\rm Coulomb}+Q_{\rm anomaly}\label{elec}
\eea
in which $Q_{\rm Coulomb}$ is the Coulomb charge
\be
\label{Coul}
Q_{\rm Coulomb}=
%\frac{1}{4\pi}
\int\pa_kF_{k0}\ d^3x=
%\frac{1}{4\pi}
\int F_{k0}\ dS^k\ ,
\ee
which can be evaluated as a surface integral by Gauss' Theorem,
and $Q_{\rm anomaly}$ is the contribution to the electric charge from the anomaly term of Ref.~\cite{Callan:1983nx},
\bea
Q_{\rm anomaly}&=&+2\ka\int\vep_{ijk}\pa_k\left[
(g-\sin g)\pa_i\cos\Ta\,(A_j-\pa_j\F)+F_{ij}\,g\,\cos\Ta\right]
\,d^3x  \ . \label{Qe0}
\eea
 
 %%%%%%%%%%%%%%%%%%%%%%%%%%%%%%%%%%%%%%%%%%
\section{Imposition of axial symmetry}
%%%%%%%%%%%%%%%%%%%%%%%%%%%%%%%%%%%%%%%%%%
Imposition of axial symmetry in the $(x_1,x_2)$ plane on
the Abelian connection $A_\mu=(A_{\bar\al},A_z,A_0)$
 (with $\bar\al=1,2$) 
in the static limit is
\be
\label{gaugecon1}
A_{\bar\al}=\left(\frac{a(\rho,z)-m}{\rho}\right)(\vep\hat x)_i\ ,\quad A_z=c(\rho,z)\ ,\quad A_0=b(\rho,z)
\ee
where $\rho=\sqrt{|x_{\bar\al}|^2}$, leading to the components of the curvature
\be
\label{gaugecurv1}
F_{\bar\al\bar\bt}=-\frac{\pa_\rho a}{\rho}\,\vep_{\bar\al\bar\bt}\ ,\quad F_{{\bar\al} z}=\pa_\rho c\ \hat x_{\bar\al}-\frac{\pa_z a}{\rho}(\vep\hat x)_{\bar\al}\ ,
\quad F_{{\bar\al} 0}=\pa_\rho b\,\hat x_{\bar\al}\ ,\quad F_{z0}=\pa_z b\,.
\ee

Imposition of symmetry on the scalar is expressed conveniently
in terms of the constraint compliant parametrisation of $\f^a=\f^a[g(x_\mu),\Ta(x_\mu),\F(\F)]$
given in \re{U} and \re{param} used above, which can be combined in the form
\be
\label{paramx}
\f^a=\left(\begin{array}{l}\f^\al\\ \f^3\\ \f^4\end{array}\right)
=\left(\begin{array}{l}\sin\frac{g(x_\mu)}{2}\sin\Ta(x_\mu)\,m^\al\\ \sin\frac{g(x_\mu)}{2} \cos\Ta(x_\mu)\\ \cos\frac{g(x_\mu)}{2}\end{array}\right)\ ,\quad{\rm with}\quad
m^\al[\F]=\left(\begin{array}{l}\cos\F(x_\mu)\\ \sin\F(x_\mu)\end{array}\right)
\ee
in which notation imposition of axial symmetry on the Skyrme scalar is stated by
\be
\label{SkSc}
g(x_\mu)=g(\rho,z)\ ,\quad\Ta(x_\mu)=\Ta(\rho,z)\ ,\quad\F(x_\mu)=m\,\vf
\ee
where $\vf$ is the azimuthal angle in the $(x_1,x_2)$ plane and $m$ is an integer.

It is useful to note that subject to this symmetry
\be
\label{useful}
(A_{\bar\al}-\pa_{\bar\al}\F)=\frac{a(\rho,z)}{\rho}\,(\vep\hat x)_{\bar\al}\ ,\quad (A_z-\pa_z\F)=c(\rho,z)
\ ,\quad (A_0-\pa_0\F)=b(\rho,z)\,.
\ee
Also, one can easily show that the magnetic potential 
$c(\rho,z)$
can be consistently set to zero, a choice we shall employ in
what follows.

%%%%%%%%%%%%%%%%%%%%%%%%%%%%%%%%%%%%%%%%%%
\subsection{Gauge deformed baryon number}
%%%%%%%%%%%%%%%%%%%%%%%%%%%%%%%%%%%%%%%%%%
The question here is whether the baryon number can be altered due to the influence of the gauge field. For this, the relevant
definition of the gauge deformed baryon number density is \re{vr21}
\[
\vr=\vr_0+\vr_1
\]
$\vr_0$ being the baryon number density prior to gauging, and $\vr_1$ given by
\be
\vr_1=\vep_{ijk}\,\pa_i(A_j\,\vep^{AB}\f^B \pa_k\f^A)\label{ro1}
\ee
which quantifies the departure of the baryon number due to gauging.

It is convenient to express these quantities in constraint compliant parametrisation given in Appendix {\bf A} in terms of the
functions $g(x_\mu)$ and $\Ta(x_\mu)$, which after imposition of axial symmetry yield
\bea
\vr_0&=&\frac{1}{2\rho}\,m\,\pa_{[\rho}(g-\sin g)\,\pa_{z]}\cos\Ta\label{redvr0}\\
\vr_1&=&-\frac{1}{2\rho}\pa_{[\rho}\left[(a-m)\,(\pa_{z]}g\,\cos\Ta+\sin g\,\pa_{z]}\cos\Ta)\right]\label{redvr1}
\eea                
in which the functions $g(\rho,z)$ and $\Ta(\rho,z)$ depend on $(\rho,z)$ subject to \re{SkSc}, as does the function $a(\rho,z)$ 
due to imposition of symmetry \re{gaugecon1}.

The integrals of \re{redvr0} and \re{redvr1} yield the deformed baryon number $q=q_0+q_1\,,$ with
\bea
q_0&=&2\pi\int\,\vr_0\,\rho\,d\rho dz=m\,\pi\int\,\pa_{[\rho}(g-\sin g)\,\pa_{z]}\cos\Ta)d\rho\,dz\label{q0}\\
q_1&=&2\pi\int\,\vr_1\rho\,d\rho dz=-\pi\int\,\pa_{[\rho}\left[(a-m)\,(\pa_{z]}g\,\cos\Ta+\sin g\,\pa_{z]}\cos\Ta)\right]d\rho\,dz\,,\label{q1}
\eea
both being double integrals with ``curl'' integrands, which are evaluated using Green's Theorem.

The integral \re{q1} vanishes since $a(\rho=0)=m$ all along the $z$-axis, while the integral \re{q0} for $q_0$ is
\be
\label{q0fin}
q_0=\pm\,m\,\pi\int\,(g-\sin g)\,\pa_{z}\cos\Ta)\big|_{\rho=0}\,dz=\pm 4m\,\pi^2\,.
\ee

The conclusion is that the baryon number cannot be changed since all Skyrmion solutions are characterised by the asymptotic condition $a(\rho=0)=m$.

%%%%%%%%%%%%%%%%%%%%%%%%%%%%%%%%%%%%%%%%%%
\subsection{Electric charge}
%%%%%%%%%%%%%%%%%%%%%%%%%%%%%%%%%%%%%%%%%%
The evaluation of the charge integral \re{Coul} for  $Q_{\rm Coulomb}$ is standard, by applying Gauss Theorm, yielding
\be
\label{QC}
Q_{\rm Coulomb}=\int\,\pa_i F_{i0}\,d^3x=4\,\pi\,(r^2\pa_r b)\big|_{r=\infty}=4\,\pi\,Q_0
\ee
where $Q_0$ is the constant in the asymptotic behaviour $b_{r\to\infty}={\rm const.}+ {Q_0}/{r^2}+\dots$.

By contrast the Gauss Theorem cannot be applied to evaluate $Q_{\rm anomaly}$, \re{Qe0}, since the function
$\Ta$ in the integrand has a discontinuity in the $(x_1,x_2)$ plane. We proceed to impose axial symmetry on the integrand $I$ of
\re{Qe0}, which yields the
%\bea
%\pa_k\om_k&=&\pa_kF_{k0}+2\ka\,\vep_{ijk}\,\pa_k\left[
%(g-\sin g)\,\pa_i\cos\Ta\,(A_j-\pa_j\F)+F_{ij}\,g\,\cos\Ta\right]\label{Qe01}\\
%\om_k&=&F_{k0}+2\ka\,\vep_{ijk}\left[
%(g-\sin g)\,\pa_i\cos\Ta\,(A_j-\pa_j\F)+F_{ij}\,g\,\cos\Ta\right]\label{Qe1}
%\eea
%
%The integrand of the volume integral
%\re{Qe01} reduces to the
following two dimensional density
\be
I=
%\frac{2\ka}{\rho}\left[a\,\pa_{[\rho}\cos\Ta\,\pa_{z]}(g-\sin g)
%(g+\sin g)\pa_{[\rho}a\pa_{z]}\cos\Ta+2\cos\Ta\,\pa_{[\rho}a\pa_{z]}g\right]\label{redQe1}\\
-\frac{2\ka}{\rho}\pa_{[\rho}\left\{cos\Ta\left[(a\pa_{z]}g-g\pa_{z]}a)-\pa_{z]}(a\,\sin g)\right]\right\}\,.\label{redQe2}
\ee
Note that after multiplying with $\rho$ coming from the volume element, the
total divergence in three dimensions reduces to a two dimensional {\bf curl}.

After integration over the azimuthal angle $\vf$, the two dimensional integral in $(\rho,z)$ is
\be
Q_{\rm anomaly}=-2\pi\,\ka\int\,\pa_{[\rho}\left\{\cos\Ta\left[(a\pa_{z]}g-g\pa_{z]}a)-\pa_{z]}(a\,\sin g)\right]\right\}\,d\rho\,dz\label{z}
\ee
that can be integrated applying Green's Theorem, yielding
\be
\label{lzz}
Q_{\rm anomaly}=-2\pi\,\ka\,m\int_{z=-\infty}^{z=+\infty}\cos\Ta\,\pa_{z}(g-\sin g)\big|_{\rho=0}\,dz
\ee
where the symmetry restriction \re{gaugecon1} stating $a(\rho=0,z)=m$ along the $z$-axis
has been imposed.

Further, imposing the boundary conditions 
employed in the numerical construction 
of the
solutions, \re{lzz} is evaluated to give              
\be
\label{lzzCW}
Q_{\rm anomaly}=4\pi\,\ka\,m
\ee
resulting in the following expression of the electric charge
\be
Q_e=4\pi(Q_0+\ka\,m)
\label{Qe}
\ee

%%%%%%%%%%%%%%%%%%%%%%%%%%%%%%%%%%%%%%%%%%
\subsection{Angular momentum}
%%%%%%%%%%%%%%%%%%%%%%%%%%%%%%%%%%%%%%%%%%
Imposition of axial symmetry \re{gaugecon1}, \re{gaugecurv1}, \re{SkSc} and \re{useful} on ${\cal J}$, \re{calJ} results in
\bea
\label{calJ1}
{\cal J}&\simeq&\left(a_{,\rho}b_{,\rho}+a_{,z}b_{,z}\right)+\nonumber\\
&&+ab\sin^2\frac g2\sin^2\Ta\left\{
%\eta^2
1+\left[
 \frac14\left(g_{,\rho}^2++g_{,z}^2\right)            
 +\left(\Ta_{,\rho}^2+\Ta_{,z}^2\right)\sin^2\frac g2
\right]\right\}
\eea
and the angular momentum is
\be
J=\int{\cal J}\,d^3x=\int{\cal J}\,d^3x=2\pi\int{\cal J}\,\rho\, d\rho dz\label{J}\,.
\ee

The quantity \re{J} can be evaluated by extracting the partial derivatives on the function $a$,% yielding
\bea
\label{J1}
J&=&2\pi\int d\rho dz\bigg(\pa_\rho(\rho\,a\,b_{,\rho})+\pa_z(\rho\,a\,b_{,z})\nonumber\\
&&\qquad\qquad\quad-a\left[\pa_\rho(\rho\,b_{,\rho})+\pa_z(z\,b_{,z})\right]\nonumber\\
&&+\rho\,ab\sin^2\frac g2\sin^2\Ta\left\{
%\eta^2
1
+\left[
 \frac14\left(g_{,\rho}^2+g_{,z}^2\right)            
 +\left(\Ta_{,\rho}^2+\Ta_{,z}^2\right)\sin^2\frac g2
\right]\right\}\bigg)
\eea
and substituting the equations of motion of the function $b$ in the second line in \re{J1}.
% as was done in \cite{VanderBij:2001nm}.

%After imposition of symmetry, the reduced two dimensional Lagrangian resulting from \re{L24} is
%\bea
%L&=&-\frac12r^2\sin\ta\bigg\{\frac{1}{r^2\sin^2\ta}\left(a_{,r}^2+\frac{1}{r^2}a_{,\ta}^2\right)
%-\left(b_{,r}^2+\frac{1}{r^2}b_{,\ta}^2\right)\nonumber\\
%&&\qquad\qquad+\eta^2\bigg[\frac14\left(g_{,r}^2+\frac{1}{r^2}g_{,\ta}^2\right)            
% +\left(\Ta_{,r}^2+\frac{1}{r^2}\Ta_{,\ta}^2\right)\sin^2\frac g2\nonumber\\
%&&\qquad\qquad\qquad\qquad+\left(\frac{a^2-b^2\,r^2\,\sin^2\ta}{r^2\,\sin^2\ta}\right)\sin^2\frac g2\sin^2\Ta
%\bigg]+\dots\bigg\}+\hat\Om_{\rm anomaly}
%\eea
%{\bf where I have omitted the quartic kinetic Skyrme term}, and $\hat\Om_{\rm anomaly}$ is the
%reduced two dimensional density resulting from the imposition of symmetry on \re{LCW}.

This analysis for the $SO(2)$ gauged Skyrme system in the absence of the anomaly term $\Om_{\rm CW}$ has been carried out 
previously in detail in Ref.~\cite{Radu:2005jp}, and since the latter does not enter the
definition of the stress-energy tensor, the details are not repeated here.

Denoting the contribution of the anomaly free part of the system to the integral \re{J1} by $J_{\rm Coulomb}$, the result is,
as per \cite{Radu:2005jp} ,
\be
\label{JC}
J_{\rm Coulomb}=4\pi\,m\,Q_0
\ee
in the notation of \re{QC}.

To calculate $J_{\rm anomaly}$, the contribution of the anomaly term to \re{J1}, the $b$
equation  resulting from the variation of $\Om_{\rm CW}$ w.r.t. $A_0$ must be employed.
This follows simply from setting the index $\tau=0$ in $\delta_{A_\tau}\tilde\Om_{\rm CW}$ equation \re{dA}.

After imposition of axial symmetry by \re{gaugecurv1}, \re{SkSc} and \re{useful}, one has
\bea
\delta_{A_0}\Om_{\rm CW}&=&\frac1\rho\left[a(\pa_{[\rho}(g-\sin g)\,\pa_{z]}\cos\Ta)
-2\,\pa_{[\rho}a(\pa_zg\cos\Ta+\sin g\,\pa_{z]}\cos\Ta)
\right]\,.\label{notcurl1}%\\
%&=&\frac{1}{r^2\sin\ta}\left[a(\pa_{[r}(g-\sin g)\,\pa_{\ta]}\cos\Ta)2\,\pa_{[r}a(\pa_zg\cos\Ta+\sin g\,\pa_{\ta]}\cos\Ta)
%\right]\,,\label{notcurl2}
\eea
leading to
\bea
J_{\rm anomaly}
&=&2\pi\int d\rho\,dz\,\pa_{[\rho}\left[a^2(\pa_{z]}g\,\cos\Ta+\sin g\,\pa_{z]}\cos\Ta)
\right]\,,\label{JCW1}
\eea
which is evaluated using Green's Theorem to give
\bea
J_{\rm anomaly}
&=&2\pi\,m^2\int_{z=-\infty}^{z=+\infty}\,(\pa_{z}g\,\cos\Ta+\sin g\,\pa_{z}\cos\Ta)\bigg|_{\rho=0}\,dz\,\,\label{JCW2}
\eea
where we have used $a(\infty)=m$, and further using $\Ta=0$ when $z<0$ and $\Ta=\pi$ at when $z>0$, finally resulting in
\be
\label{JCWfin}
J_{\rm anomaly}=4\pi\,m^2\,.
\ee

The total angular momentum $J=J_{\rm Coulomb}+\ka\,J_{\rm anomaly}$ is finally
\be    
J=4\pi\,m(Q_0+\ka\,m)=m\,Q_e\,.
\ee

%%%%%%%%%%%%%%%%%%%%%%%%%%%%%%%%%%%%%%%%%%
\section{The solutions}
%%%%%%%%%%%%%%%%%%%%%%%%%%%%%%%%%%%%%%%%%%

%%%%%%%%%%%%%%%%%%%%%%%%%%%%%%%%%%%%%%%%%%
\subsection{The Ansatz, effective action
 and boundary solutions}
%%%%%%%%%%%%%%%%%%%%%%%%%%%%%%%%%%%%%%%%%%

The numerics is done in spherical coordinates
$(r,\theta,\varphi)$,
by employing an Ansatz 
 in terms of five functions
\begin{eqnarray} 
 \{ \Phi_1(r,\theta),~\Phi_2(r,\theta),~\Phi_3(r,\theta) \} 
~~~{\rm with}~~~ \Phi_1^2+\Phi_2^2+\Phi_3^2=1,~~{\rm and}~~
~a (r,\theta) ,~b(r,\theta)  , 
\end{eqnarray}
  the three real scalars
	$\Phi_a$
	being
related to 
the fields $\{ \phi^I ,\phi^4 \}$ in the Skyrme field Ansatz~\eqref{fU}, 
as follows
\begin{eqnarray}
\label{Uans}
 \phi^1+i \phi^2=\Phi_1(r,\theta)e^{i ( m\varphi -\om t) }\ , \qquad
 \phi^3=\Phi_2(r,\theta) \ , \qquad \phi^4 =\Phi_3(r,\theta) \ , 
\end{eqnarray}
where  $m$ an integer.
 Also, to make contact with the previous work in the literature
on the gauge decoupling limit,
and to make more transparent the local U(1) symmetry
of the reduced action,
we have introduced the harmonic frequency $\om$ and the winding 
number $m$ in the Skyrme field Ansatz,
  together with the gauge coupling constant $e$
in the covariant derivative.
%
% we have introduced the harmonic frequency $\om$
 %in the Skyrme field Ansatz
%(although we shall report results for  
%$\om=0$).
% 
%The functions
%$a$ and $b$
%are the magnetic and electric potentials in
The gauge field Ansatz is 
\begin{eqnarray}
\label{An} 
 A_\mu dx^\mu=a_\varphi (r,\theta) d\varphi+b(r,\theta) dt~,
\end{eqnarray} 
with $a_\varphi$
and $b$ the  magnetic and electric potentials, respectively
(with $a_\varphi=a-m$ in (\ref{gaugecon1})).

This  axially symmetric Ansatz  
results in the following  two-dimensional effective action 
\begin{eqnarray}
S_{eff}= 
%\frac{\pi  F_\pi^4 a^2}{8}
\int_0^\infty dr \int_0^\pi d\theta L_{eff}~,
\end{eqnarray}
with  
\begin{eqnarray}
\label{Leff}
L_{eff}= 
r^2 \sin \theta
\left(
\frac{1}{4} {\cal F}^2
+\lambda_1 \frac{1}{2}L_{2}^{(S)}
+\lambda_2 \frac{1}{4} L_{4}^{(S)}
+\mu^2 (1-\Phi_3)
\right)+\kappa L_{CW},
\end{eqnarray}
where 
\begin{eqnarray}
&&
{\cal F}^2=  
(a_{\varphi ,r}^2+\frac{a_{\varphi ,\theta}^2}{r^2})
-
(b_{,r}^2+\frac{b_{,\theta}^2}{r^2}),
\\
&&
L_{2}^{(S)}=
  \Phi_{1,r}^2+ \Phi_{2,r}^2+ \Phi_{3,r}^2
+\frac{1}{r^2}
(
 \Phi_{1,\theta}^2+ \Phi_{2,\theta}^2+ \Phi_{3,\theta}^2
)
+\phi_1^2
\left(
\frac{(m+e a_{\varphi})^2}{r^2\sin^2 \theta}
-(\om-e b)^2
\right),
\\
&&
L_{4}^{(S)}=\frac{4}{r^2}
\bigg[
(\Phi_{3,\theta} \Phi_{2,r} -\Phi_{2,\theta} \Phi_{3,r} )^2
+(\Phi_{2,\theta} \Phi_{1,r} -\Phi_{1,\theta} \Phi_{2,r} )^2
+(\Phi_{3,\theta} \Phi_{1,r} -\Phi_{1,\theta} \Phi_{3,r} )^2
\\
\nonumber
&&
{~~~~~~~~~}
+r^2 \phi_1^2
\left(
\frac{(m+e a_{\varphi})^2}{r^2 \sin^2 \theta}
- (\om-e b)^2
\right)
\left(
 \Phi_{1,r}^2+\Phi_{2,r}^2+\Phi_{3,r}^2
+\frac{1}{r^2} (\Phi_{1,\theta}^2+\Phi_{2,\theta}^2+\Phi_{3,\theta}^2)
\right)
\bigg],
\end{eqnarray}
while the CW term reads 
\begin{eqnarray}
 &&
\nonumber
L_{CW}=
2\Phi_1 
        \left(\om (a_{\varphi}+\frac{m}{e})+m (b-\frac{\om}{e})  \right)
				\\
				\nonumber
&&
{~~~~~~~~~~}
\bigg(
 \Phi_1( \Phi_{2,\theta}  \Phi_{3,r}- \Phi_{3,\theta}  \Phi_{2,r})
+\Phi_2( \Phi_{3,\theta}  \Phi_{1,r}- \Phi_{1,\theta}  \Phi_{3,r})
+\Phi_3( \Phi_{1,\theta}  \Phi_{2,r}- \Phi_{2,\theta}  \Phi_{1,r})
\bigg)
\\
 &&
	{~~~~~~~~~~}
e \bigg[
 \Phi_2   \left(
(b-\frac{\om}{e}) (a_{ \varphi,\theta}\Phi_{3,r}-a_{\varphi ,r}\Phi_{3, \theta} )
+(a_{\varphi}+\frac{m}{e}) b_{ ,r}\Phi_{3,\theta}-b_{ ,\theta}\Phi_{3,r} )
   \right),
	\\
	\nonumber
	&&
	{~~~~~~~~~~}
 +\Phi_3   \left(
(b-\frac{\om}{e}) (a_{\varphi,r}\Phi_{2,\theta}-a_{\varphi, \theta}\Phi_{2, r} )
+(a_{\varphi}+\frac{m}{e})(b_{ ,\theta}\Phi_{2,r}-b_{ ,r}\Phi_{2,\theta} )
   \right)	
\bigg]
	\\
	\nonumber
	&&
	{~~~~~~}
	+ 
 (\Phi_3 \Phi_{2,\theta}-\Phi_2 \Phi_{3,\theta})( \om a_{\varphi ,r}+ m b_{,r})
+(\Phi_2 \Phi_{3,r}-\Phi_3 \Phi_{2,r})( \om a_{\varphi ,\theta}+ m b_{,\theta}).
\end{eqnarray}
The constant
$ \lambda_1, \lambda_2$,
$\lambda_M$,
$\mu$
and 
$\kappa$
in the above relations 
are input parameters
(with $\mu$  corresponding to the mass term for the Skyrme field).
% related to those defined in
%(\ref{action}).
 The main reason to introduce 
 these constants  (in particular 
 $\lambda_2$,
 $\mu$
 and $\kappa$)
 is to keep track of several limits of interest.
%of the contribution of various individual terms
%to various quantities of interest. 
 The precise values used in the numerics will be specified in Section {\bf 5.2} (and they will be chosen to match Eqs.~(\ref{U1Sk1}) and (\ref{U1Sk2})).

Also, let us remark the existence of a residual gauge symmetry,
with the  gauge field potentials always appearing in the combination
\begin{eqnarray}
 (\om-e b)~~{\rm and}~~~(m+e a_\varphi)~ .
\end{eqnarray} 
This symmetry is fixed by imposing 
$\om=0$,
and the magnetic
 gauge potential  to vanish at infinity.

%%%%%%%%%%%%%%%%%%%%%%%%%%%%%%%%%%%%%%%%%%%%%%%%%%%%%%%%%%%%%%%%%%%%%%%%%%%%%%%%%%%%%%
%\subsection{Boundary conditions}
%%%%%%%%%%%%%%%%%%%%%%%%%%%%%%%%%%%%%%%%%%%%%%%%%%%%%%%%%%%%%%%%%%%%%%%%%%%%%%%%%%%%%%

The solutions are found numerically, by solving a boundary value problem.
The imposed boundary conditions are
\begin{eqnarray} 
&&
\Phi_1 \big|_{r=0}=0,~
\Phi_2 \big|_{r=0}=0,~
\Phi_3 \big|_{r=0}=-1,~ 
a_\varphi  \big|_{r=0}=0,~ 
\partial_r b \big|_{r=0}=0,~ 
\\
&&
\Phi_1 \big|_{r= \infty}=0,~
\Phi_2 \big|_{r= \infty}=0,~
\Phi_3 \big|_{r= \infty}=1,~ 
a_\varphi  \big|_{r= \infty}=0,~ 
b \big|_{r= \infty}=V,~ 
\end{eqnarray}
 with $V$
a constant corresponding to the electrostatic potential at infinity 
(which should not be confused with the potential term in (\ref{Lin})),
and
\begin{eqnarray}  
\Phi_1 \big|_{\theta=0,\pi}=0,~
\partial_\theta \Phi_2 \big|_{\theta=0,\pi}=0,~
\partial_\theta \Phi_3 \big|_{\theta=0,\pi}=0,~
a_\varphi  \big|_{\theta=0,\pi}=0,~
\partial_\theta b  \big|_{\theta=0,\pi}=0.
\end{eqnarray}
The solutions also possess a 
reflexion symmetry $w.r.t.$ the equatorial plane,
where we impose
\begin{eqnarray}  
\partial_\theta \Phi_1 \big|_{\theta=\pi/2}=0,~
\Phi_2 \big|_{\theta=\pi/2}=0,~
\partial_\theta \Phi_3 \big|_{\theta=\pi/2}=0,~
\partial_\theta a_\varphi  \big|_{\theta=\pi/2}=0,~
\partial_\theta b \big|_{\theta=\pi/2}=0.
\end{eqnarray}

%%%%%%%%%%%%%%%%%%%%%%%%%%%%%%%%%%%%%%%%%%%%%%%%%%%%%%%%%%%%%%%%%%%%%%%%%%%%%%%%%%%%%%
 %\subsection{Imposing the sigma-model constraint}
%%%%%%%%%%%%%%%%%%%%%%%%%%%%%%%%%%%%%%%%%%%%%%%%%%%%%%%%%%%%%%%%%%%%%%%%%%%%%%%%%%%%%%

% 
In our approach
the sigma-model constraint is not imposed directly,
but rather emerges within the numerical scheme,
where we use the
Lagrange multiplier method \cite{Rajaraman:1982is}.
In this approach,
the effective Lagrangian of the model is supplemented with the constraint
\begin{eqnarray}
L_{eff} \to L_{eff}^{(0)}+ \sqrt{-g} \left(\sum_a \Phi^a \Phi^a-1 \right){\cal C},
\end{eqnarray}
with $L_{eff}^{(0)}$ the initial effective lagrangian 
(\ref{Leff})
and ${\cal C}$ is {\it a new function}.
Then the  equations 
for $\Phi_a$
get an extra-contribution,
\begin{eqnarray} 
\label{eqa}
\mathbb{E}q_a = \mathbb{E}q_a^{(0)}+2 {\cal C} \Phi_a \sqrt{-g} =0,~~
{\rm with}~~
\mathbb{E}q_a^{(0)}=
\partial_r \left(
\frac{\partial L_{eff}  }{\partial(\partial_r \Phi_a)} 
           \right)
      +\partial_\theta
			\left( 
			\frac{ L_{eff}}{\partial
(\partial_\theta \Phi_a)}
           \right)
-\frac{\partial  L_{eff}}{\partial \Phi_a},
\end{eqnarray}
while the variation $w.r.t.$
${\cal C}$ imposes the sigma-model  contraint as an extra equation.
Then the Eqs. (\ref{eqa}) are multiplied each one with $\Phi_a$,
and one takes  their sum.
After using also the equation for ${\cal C}$, one finds 
\begin{eqnarray} 
\label{fxz}
{\cal C}=-\frac{1}{2 \sqrt{-g} } \Phi_a (\mathbb{E}q_a^{(0)}).
\end{eqnarray}
Thus, in the numerics  
the eqs. for the the Skyrme fields 
$\Phi_a$
 are still 
 (\ref{eqa}),
 with ${\cal C}$
given by (\ref{fxz}).

All numerical calculations in the axially symmetric case have been
performed by using a professional package,
which uses a  Newton-Raphson finite difference method with
an arbitrary grid and arbitrary consistency order \cite{schoen}. 
In numerics, a compactified radial coordinate $x$ was introduced, with
\begin{eqnarray}  
x=\frac{r}{1+r},
\end{eqnarray}
 the  equations being discretized on a ($x,~\theta$) grid with around $150\times 50$ points.
Then the resulting system is solved iteratively until convergence is achieved.
The typical  numerical error
for the solutions reported in this work is estimated to be of the order of $10^{-3}$
(also, the order of the difference formulae  was six).

%%%%%%%%%%%%%%%%%%%%%%%%%%%%%%%%%%%%%%%%%% 
\begin{figure}[ht!]
%\lbfig{rhfar}
\begin{center}
\includegraphics[height=.34\textwidth, angle =0 ]{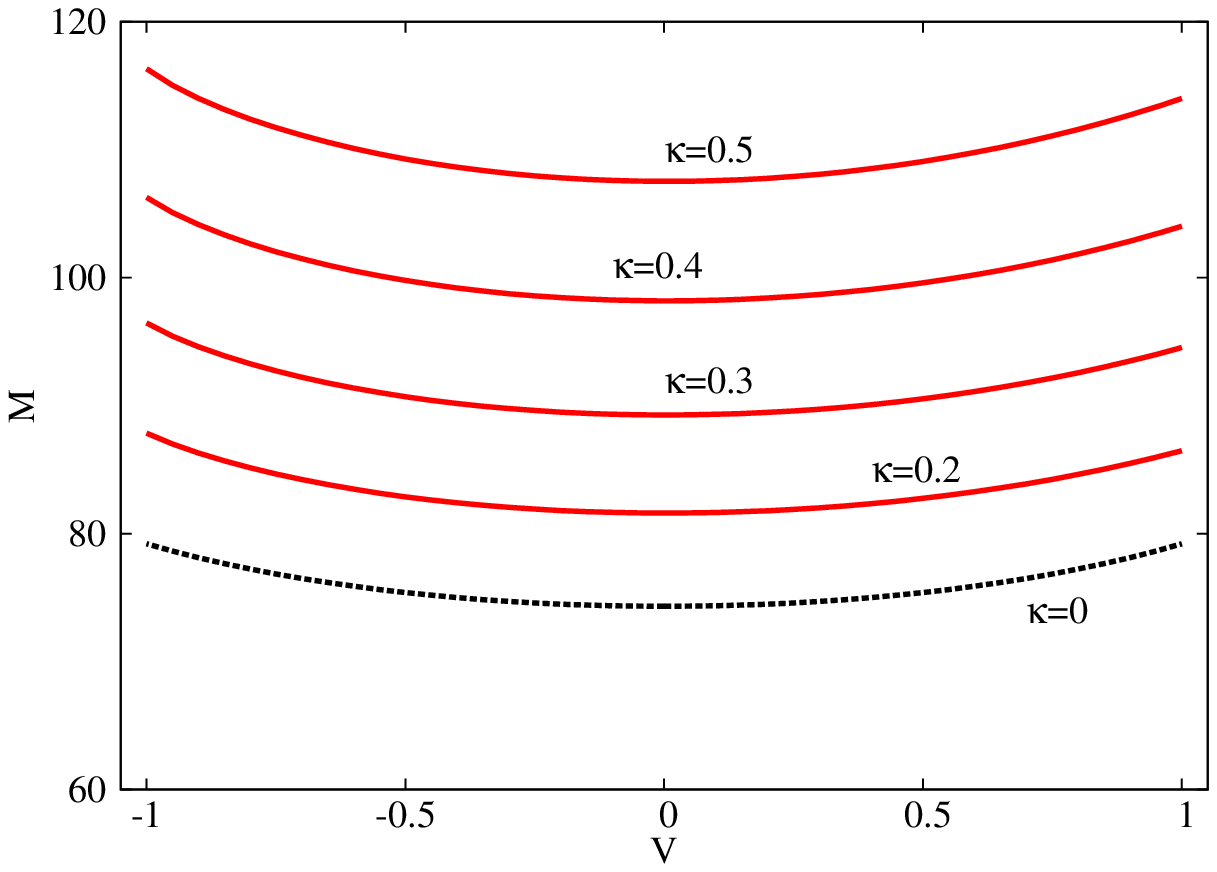}
\includegraphics[height=.34\textwidth, angle =0 ]{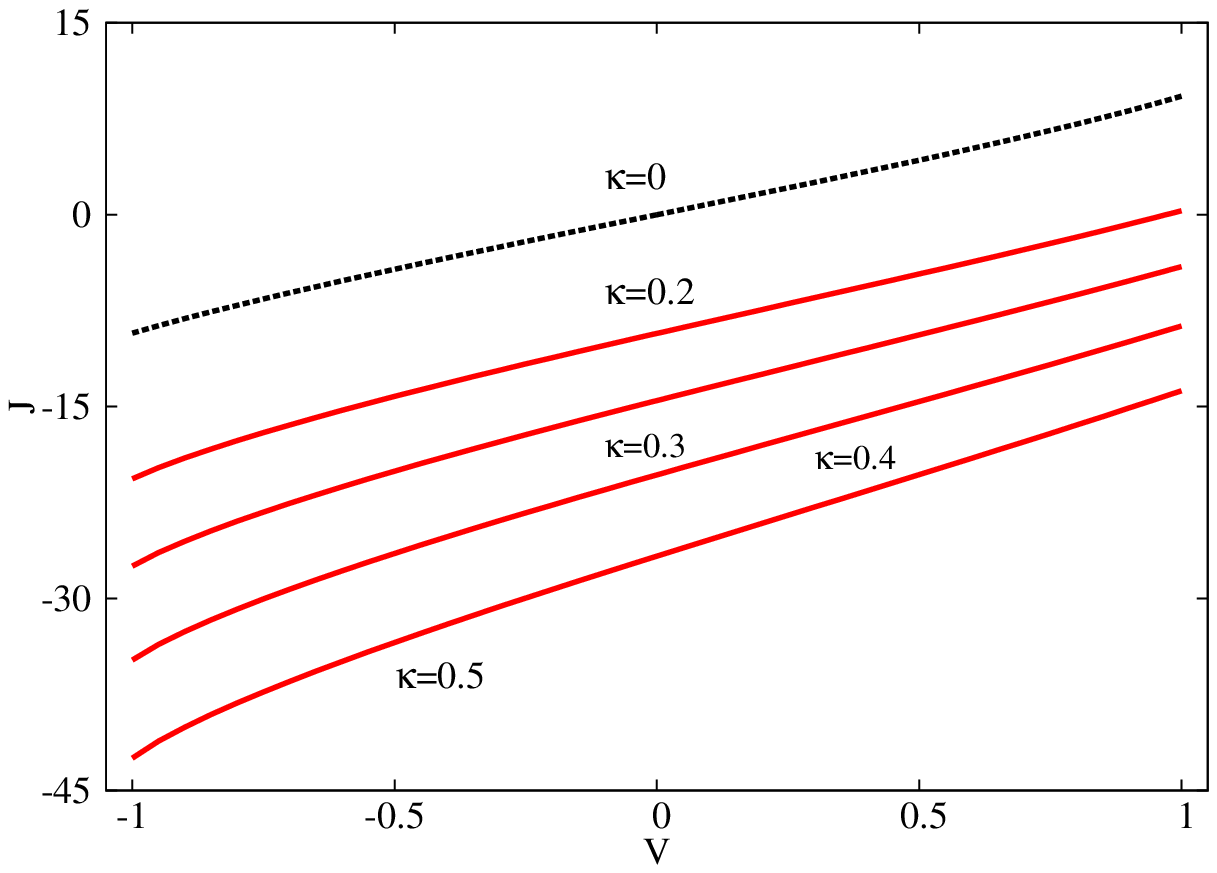}
\end{center}
\caption{
 Mass (left panel) and angular momentum (right panel) for
gauged Skyrmions are shown in terms of electrostatic potential at infinity
 $V$  
for several values of the constant $\kappa$
multiplying the Callan-Witten term.
}
\label{fig1}
\end{figure}
%%%%%%%%%%%%%%%%%%%%%%%%%%%%%%%%%%%%%%%%%%% 
%%%%%%%%%%%%%%%%%%%%%%%%%%%%%%%%%%%%%%%%%% 
\begin{figure}[ht!]
%\lbfig{rhfar}
\begin{center}
\includegraphics[height=.34\textwidth, angle =0 ]{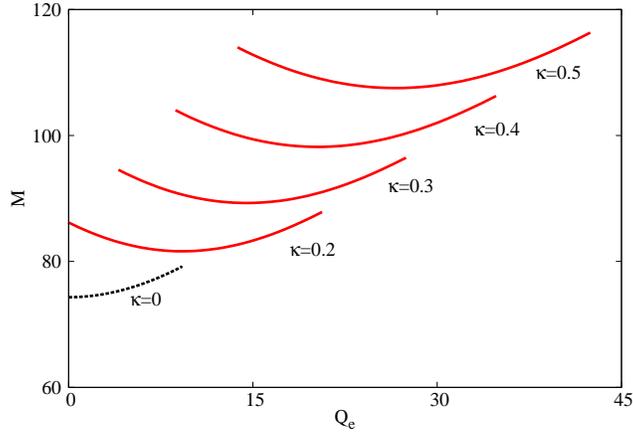}
\end{center}
\caption{
%{\it Left panel}:
The mass of gauged Skyrmions is shown as a function of electric charge 
 for several values of the input parameter
 $\kappa$.
}
\label{fig2}
\end{figure}
%%%%%%%%%%%%%%%%%%%%%%%%%%%%%%%%%%%%%%%%%%%

The total mass-energy and angular momentum
 of the configurations is defined as the volume integral over all space
of the 
$T_{0}^0$ 
and 
$T_{\varphi}^0$
components of the stress-energy tensor, 
with
$
M = \int d^3x T_{0}^0 
$
and
$
J= \int d^3x T_{\varphi}^0.
$
In addition, the  configurations
possess also a nonzero magnetic dipole moment $\mu_{mag}$ \cite{Piette:1997ny},
which is read from the asymptotics of
the magnetic potential, 
$a_\varphi \to {\mu_{mag}\sin^2\theta}/{r}+\dots$.

%%%%%%%%%%%%%%%%%%%%%%%%%%%%%%%%%%%%%%%%%%%%%%%%%%%%%%%%%%%%%%%%
\begin{figure}[ht!]
%\lbfig{rhfar}
\begin{center}
 \includegraphics[height=.34\textwidth, angle =0 ]{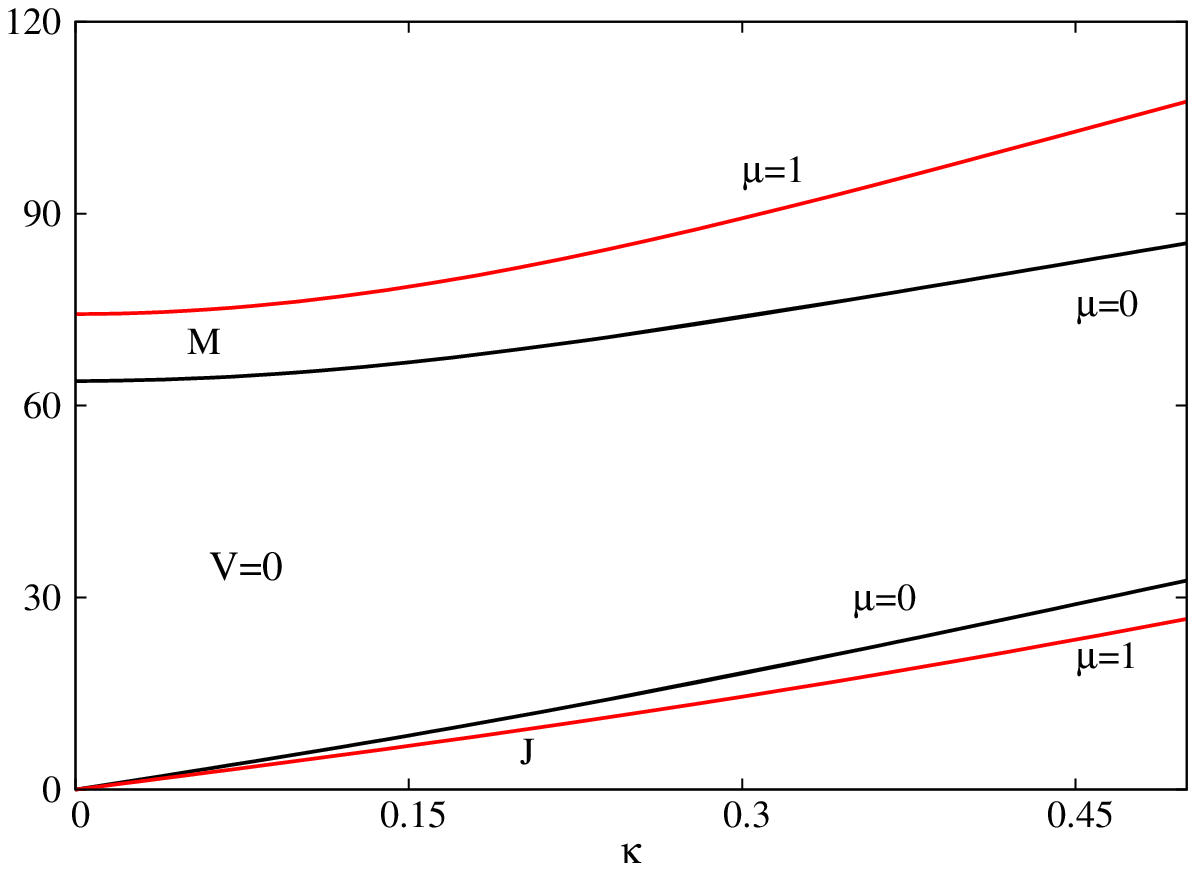}
\includegraphics[height=.34\textwidth, angle =0 ]{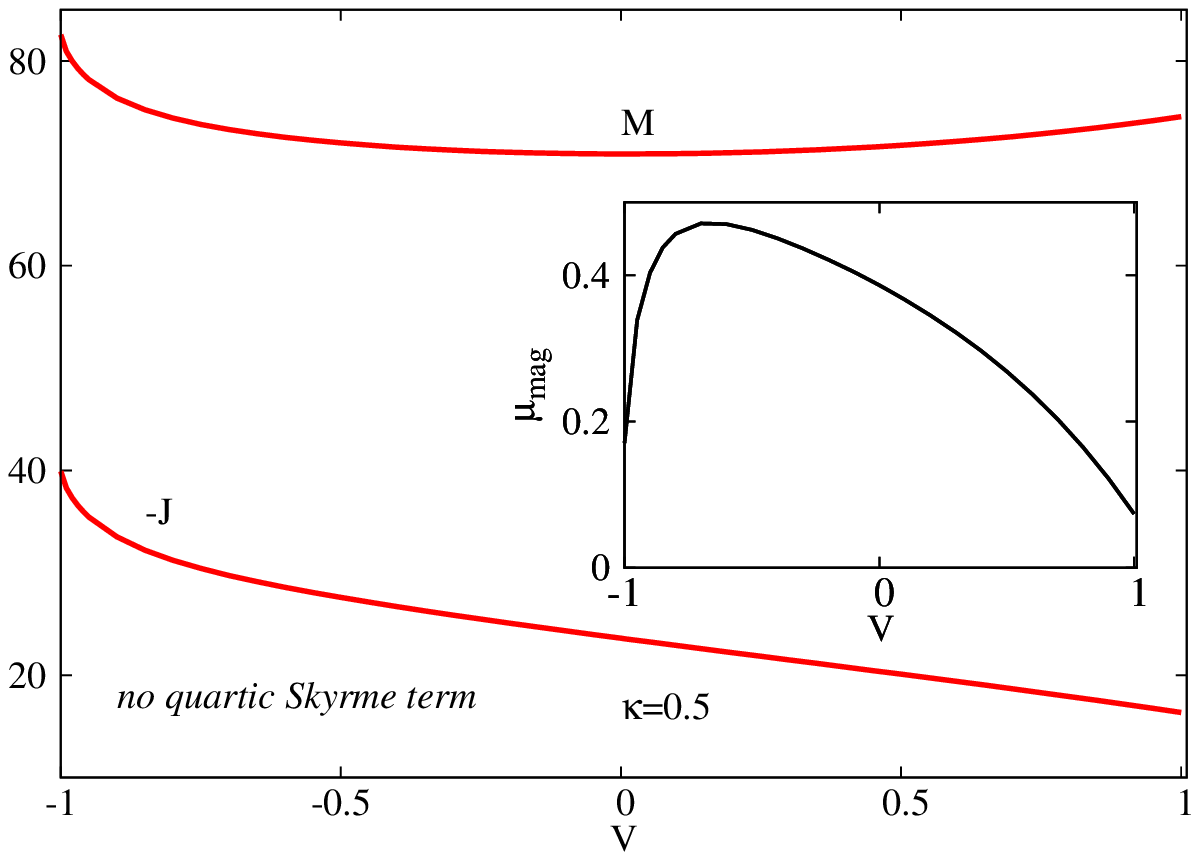}
\end{center}
\caption{
{\it Left panel:}
 The mass and angular momentum are shown
as a function of the constant $\kappa$  
 for the special 
solutions with a vanishing
electric potential at infinity,
and two values of the pion mass $\mu$. 
{\it Right panel:}
 Mass, angular momentum and magnetic dipole moment
are shown in terms of electrostatic potential at infinity
 $V$  
 for gauged solutions in a model without a quartic Skyrme term.
}
\label{fig3}
\end{figure}
%%%%%%%%%%%%%%%%%%%%%%%%%%%%%%%%%%%%%%%%%%%

%%%%%%%%%%%%%%%%%%%%%%%%%%%%%%%%%%%%%%%%%%%% 
\subsection{Numerical results}
%%%%%%%%%%%%%%%%%%%%%%%%%%%%%%%%%%%%%%%%%%%% 

Since the model is invariant under the change of sign of
$A_0\equiv b$ and $\kappa$,
we shall restrict our study to the case 
$\kappa\geq 0$.
Also, all numerics is done for 
unit winding number, $m=1$,
$i.e.$ a unit topological charge.
Moreover, the value of the gauge coupling constant is set to one.
  
Concerning the choice of other parameters in 
the reduced Lagrangian (\ref{Leff}),
  the numerics is done for the choice
\begin{eqnarray}
\mu=\lambda_1=2\lambda_2=\lambda_M=e=1,
\end{eqnarray}
unless specified otherwise.
Also,  we report results for  several values 
of $\kappa$
ranging from zero to 1/2 (note that
solutions are likely to exist for arbitrary  
 $\kappa$; however,
the numerical errors increase
for larger $\kappa$).

\medskip

Some of the properties of the solutions
are similar to those found in the
literature for $\kappa=0$.
The solutions exist for a finite range of
the electrostatic potential at infinity,
\begin{eqnarray}  
e^2 V^2 \leq \mu^2,
%~~{\rm with}~~\mu=\lambda_0^{1/2}.
\end{eqnarray}
as found from a study of the far field behaviour of the scalar functions.
We   also  mention  that the limiting configuration with $e V = \pm \mu$
are still localized, with
 finite global charges.
The solutions carry nonzero angular momentum densities,
while
their intrinsic shape 
(as found $e.g.$ from the study of surfaces of
constant energy density) corresponds to deformed spheres.

 \medskip

Other features are new, being found for $\kappa\neq 0$ only.
First, 
as seen in Fig. \ref{fig1}
the symmetry of solutions   under the change of
sign  
$V \to -V$
(or $ Q_e \to-Q_e$) 
is lost in the presence of a CW term.
Moreover, configurations
with $J=0$
and still possessing a 
non-zero angular momentum density exist as well.

Also, 
for a fixed value 
 of the electrostatic potential at infinity,
the mass, angular momentum  (and thus also the electric charge) 
increase with $\kappa$.
At the same time, for a given model with some
 $\kappa>0$, we notice a non-monotonic behaviour of the mass of the solutions
as a function of electric charge,
with a minimal value of $M$
for some critical $Q_e$, see Fig. \ref{fig2}.

For $\kappa=0$,
the solutions with a vanishing electric potential at infinity
are static, while the electric field vanishes, $b\equiv 0$.
This no longer is the case for a nonzero $\kappa$,
where we have found
spinning, electrically charged configurations  
with $V=0$,  see Fig. \ref{fig3} (left panel).
Therefore, in this case the presence of a pion-mass term in the model
is not necessary, and configurations with
finite charges exist for
$\mu=0$ as well.

Another specific feature consists in the
existence of solutions
with finite mass and angular momentum
 in the absence of a quartic Skyrme term in the Lagrangian
(\ref{Leff}),
%\begin{eqnarray}  
$
\lambda_2=0.
$
%\end{eqnarray}
That is, such configurations 
are supported by the 
 contribution  of the CW-term,
which provides the supplementary 
contribution to fulfill the Derrick-type scaling requirement. 
 Otherwise, the main properties of these solutions
are similar to those found 
for configurations in the full model,
see Figure \ref{fig3} (right panel).
Note that, as expected, for a given $V$,
the mass and angular momentum of the $\lambda_2=0$ 
solutions is smaller 
than the corresponding solutions with a quartic term.

%%%%%%%%%%%%%%%%%%%%%%%%%%%%%%%%%%%%%%%%%%%%%%%%%%%%%%%%%%%%%%%%%%%%%%%%%%%%%%%%%%%%%5
\section{Conclusions}
%%%%%%%%%%%%%%%%%%%%%%%%%%%%%%%%%%%%%%%%%%%%%%%%%%%%%%%%%%%%%%%%%%%%%%%%%%%%%%%%%%%%%%5

The main purpose of this work was to present a preliminary investigation of
U(1)-gauged Skyrmions in $3+1$ dimensions, in a model featuring the Callan-Witten (CW) anomaly term.
To our knowledge such results have not appearred in the literature\footnote{
The CW term has been employed previously in the study \cite{Moss:2000hf}
of charged black hole solutions with pion hair.
However, the results there are found for a 
non-topological Ansatz for the Skyrme field.} to date. The main thrust of the work is to reveal the qualitative
similarity of the dynamical effects of the CW density on the gauged Skyrmion, to that of the
Chern-Simons (CS) density observed in Abelian gauged Skyrmions (in odd dimensions).

It turns out that the presence of the CW term in the Abelian gauged Skyrme system in $3+1$ dimensions has
most of the dynamical effects that the CS term has on the Abelian gauged Skyrme systems in $2+1$ and $4+1$ 
dimensions, observed previously in Refs.~\cite{Navarro-Lerida:2016omj,Navarro-Lerida:2018siw}
and \cite{Navarro-Lerida:2020jft,Navarro-Lerida:2020hph} respectively. Speifically, these effects are the
unusual dependence of the mass/energy $M$ on the electric charge $Q_e$ and angular momentum $J$, contrasting with the usual
monotonically increasing behaviour of $M$ on $(Q_e,J)$ such that $M$ displays a minimum for a critical value of $(Q_e,J)$.
In other words, the slope of $M$ $vs.$ $(Q_e ,J)$ can be positive and negative. This is the salient message here.

The exception is the effect of CW term (in $3+1$ dimensions) on the `baryon number'' $q$ as opposed to the effect of the
CS term (in $2+1$ and $4+1$ dimensions) on $q$, which differs for different solutions characterised by $a_\infty$. In the
case of the CW term, $q$ does not change. This may be a feature of Abelian gauging here, and it is conceivable that the 
situation may be different for non-Abelian gauged models, $e.g.,$ in \cite{Brihaye:2001qt,Brihaye:2002nz}.

Another effect of the CW term in this model is that $SO(2)$ gauged Skyrmion persists even when the quartic kinetic term
of the Skyrme scalar is suppressed, which contrasts with the usual case when the CW term is absent.

Finally it should be remarked that the CW density differs qualitatively from the CS density in that
its definition includes the Skyrme scalar in addition to the gauge field while the CS density is defined
exclusively in terms of the gauge field. This results in the definition of electric charge (and angular
momentum) as a Noether charge. In this respect, the CW density is akin to the Skyrme--Chern-Simons (SCS)
densities discussed most recently in Ref.~\cite{Tchrakian:2021xzy}. In this context, it may be worth remarking
that a SCS density can be defined also in odd spatial dimensions, and in particular in $2+1$ dimensions we have verified
that the salient effects of that SCS term are the same as those resulting from the CS term. This work is in
active development will appear in the near future.

All numerical results reported in this work
are for
unit winding number, $m=1$.
However, similar (axially symmetric) solutions should exist for arbitrary $m>1$.
Moreover, it would be interesting to extend these results 
for non-axisymmetric  configurations which are known to exist in the ungauged case.

%%%%%%%%%%%%%%%%%%%%%%%%%%%%%%%%%
\section*{Acknowledgements}
%%%%%%%%%%%%%%%%%%%%%%%%%%%%%%%%%
Our thanks go to Valery Rubakov (R.I.P.) for his generous  and helpful discussions. The work of E.R.
is supported  by the  Center for Research and Development in Mathematics and Applications (CIDMA) through the Portuguese Foundation for Science and Technology (FCT -- Fundac\~ao para a Ci\^encia e a Tecnologia), references  UIDB/04106/2020 and UIDP/04106/2020.  
E.R. also acknowledge support  from the projects CERN/FIS-PAR/0027/2019,
 PTDC/FIS-AST/3041/2020,  
CERN/FIS-PAR/0024/2021 and 2022.04560.PTDC.  
This work has further been supported by  the  European  Union's  Horizon  2020  research  and  innovation  (RISE) programme H2020-MSCA-RISE-2017 Grant No.~FunFiCO-777740 and by the European Horizon Europe staff exchange (SE) programme HORIZON-MSCA-2021-SE-01 Grant No.~NewFunFiCO-101086251. F.N.-L. gratefully acknowledges support  from MICINN under project PID2021-125617NB-I00 ``QuasiMode" and also from Santander-UCM under project PR44/21‐29910.

%%%%%%%%%%%%%%%%%%%%%%%%%%%%%%%%%
%\newpage 
\appendix
\setcounter{equation}{0}
\renewcommand{\theequation}{A.\arabic{equation}}
\section{The $3+1$ constraint compliant parametrisation}
%%%%%%%%%%%%%%%%%%%%%%%%%%%%%%%%%

%\subsection{A computation of the Anomaly-term expression}

The constraint-compliant parametrisation employed here is
\be
\label{U}
U=\cos\frac{g}{2}\eins+i\,m^I\si_I\sin\frac{g}{2}\ ,\quad I=1,2,3\ ;\quad{\rm and}\quad |m^I|^2=1\,,
\ee
in which according to the gauging prescription \re{coval}-\re{covA}, the function $g$ is gauge-inert and only the $m^I$
gauge-rotate.

%{\bf Indeed it is instructive to display the transformation of $m^i$ under the infinitesimal $SO(2)$
%rotation $\Delta$:
%\bea
%&&m^1\to m^1-\Delta\,m^2\label{m1}\\
%&&m^2\to m^2+\Delta\,m^1\label{m2}\\
%&&m^3\to\ m^3\label{m3}
%\eea
%where the component $m^3$ is acually gauge inert.}

In the parametrisation \re{U},
\bea
U^{-1}\pa_\mu U&=&X_\mu^K\,\si_K\label{U-du}\\
U\pa_\mu U^{-1}&=&Y_\mu^K\,\si_K\ ,\quad{\rm or}\quad \pa_\mu UU^{-1}=-Y_\mu^K\,\si_K\label{Udu-+}
\eea
where
\bea
X_\mu^K&=&\ \frac{i}{2}\left[
\pa_\mu g\,m^K+\sin g\,\pa_\mu m^K+(1-\cos g)\vep^{IJK}m_I\pa_\mu m_J
\right]\si_K\label{Xmuk}\\
Y_\mu^K&=&-\frac{i}{2}\left[
\pa_\mu g\,m^K+\sin g\,\pa_\mu m^K-(1-\cos g)\vep^{IJK}m_I\pa_\mu m_J
\right]\si_K\,,\label{Ymuk}
\eea
and we further employ the constraint compliant polar parametrisation of $m^I=(m^\al,m^3)$
\be
\label{param}
m^I=\left(\begin{array}{l}\sin\Ta(x_\mu)\cos\F(x_\mu)\\ \sin\Ta(x_\mu)\sin\F(x_\mu)\\\ \cos\Ta(x_\mu)\end{array}\right)
\ee
in which the function $\Ta(x_\mu)$ is gauge inert and the function $\F(x_\mu)$ translates
with the $SO(2)$ gauge transformation.

The quadratic kinetic Skyrme term in this parametrisation
takes the form
\be
\label{quadSk}
|D_\mu\f^a|^2=\frac14|\pa_\mu g|^2+\frac12(1-\cos g)\left[\pa_\mu\Ta|^2+\sin^2\Ta|(A_\mu-\pa_\mu\F)|^2\right]
\ee 
and the quartic kinetic Skyrme term in this parametrisation
takes the form
\bea
\label{quartSk}
&&|D_{[\mu}\f^aD_{\nu]}\f^b|^2=\frac14(1-\cos g)|\pa_{[\mu}g\,\pa_{\nu]}\Ta|^2\nonumber\\
&&+\frac12(1-\cos g)\sin^2\Ta\left[\frac12|\pa_{[\mu}g\,(A_{\nu]}-\pa_{\nu]}\F)|^2
+(1-\cos g)\pa_{[\mu}\Ta\,|(A_{\nu]}-\pa_{\nu]}\F)|^2\right]
\eea
which are both gauge invariant as they stand, since the Maxwell connection $A_\mu$ appears only in the form
\[
A_\mu-\pa_\mu\F\ .
\]

Next we calculate the CW density, to which end we start by
evaluating the traces
\bea
\vep^{\tau\la\mu\nu}\mbox{Tr}\,Q(U^{-1}\pa_\la U)(U^{-1}\pa_\mu U)(U^{-1}\pa_\nu U)&=&\ 
\frac{1}{4!}\vep^{\tau\la\mu\nu}\vep_{IJK}\,X_\la^K\,X_\mu^I\,X_\nu^J\label{TrXXX}\\
\vep^{\tau\la\mu\nu}\mbox{Tr}\,Q(U\pa_\la U^{-1})(U\pa_\mu U^{-1})(U\pa_\nu U^{-1})&=&
-\frac{1}{4!}\vep^{\tau\la\mu\nu}\vep_{IJK}\,Y_\la^K\,Y_\mu^I\,Y_\nu^J~.
\label{TrYYY}
\eea

Substituting \re{TrXXX}-\re{TrYYY} in \re{W1} yields the expression for $W_{(1)}^{\tau}$
\bea
W_{(1)}^{\tau}&=&\frac12\,\pa_\la(g-\sin g)\,
\vep^{\tau\la\mu\nu}\,\vep_{IJK}\,m^I\pa_\mu m^J\pa_\nu m^K\label{W10a}\label{W10aa}\,.
\eea

Substituting \re{Xmuk}-\re{Ymuk} in  \re{W2}, taking account of \re{term},
it follows that
\be
W_{(2)}^{\tau\mu\nu}=-
\frac{1}{2}\,\vep^{\tau\la\mu\nu}(m^3\ \pa_\la g +\sin g\ \pa_\la m^3)\label{W10bb}
\ee

In the parametrisation \re{param}, \re{W10a} and \re{W10bb} are expressed as
\bea
W_{(1)}^{\tau}
&=&2\,\vep^{\tau\la\mu\nu}\,\pa_\la(g-\sin g)\,\pa_\mu\cos\Ta\,\pa_\nu\F\label{W10ac}\\
W_{(2)}^{\tau\mu\nu}&=&
\vep^{\tau\la\mu\nu}(\pa_\la g\,\cos\Ta+\sin g\,\pa_\la\cos\Ta)\label{W10bc}
\eea
leading to the action densities
\bea
\Om_{\rm CW}^{(1)}&=&
2\,\vep^{\tau\la\mu\nu}\,A_\tau\,\pa_\la(g-\sin g)\,\pa_\mu\cos\Ta\,\pa_\nu\F\label{Om1int}\\
&=&-\vep^{\tau\la\mu\nu}\,F_{\mu\nu}\,(g-\sin g)\pa_\tau\cos\Ta\,\pa_\la\F\label{Om1eff}\\
\Om_{\rm CW}^{(2)}&=&\vep^{\tau\la\mu\nu}\,A_\tau\,F_{\mu\nu}\,(\pa_\la g\,\cos\Ta+\sin g\,\pa_\la\cos\Ta)
\label{Om2int}
\eea
where \re{Om1eff} is equivalent to \re{Om1int} up to a total divergence term., $i.e.,$
\bea
\Om_{\rm CW}&=&\vep^{\tau\la\mu\nu}\,F_{\mu\nu}\bigg[-(g-\sin g)\pa_\tau\cos\Ta\,\pa_\la\F\nonumber\\
&&\qquad\qquad\ \ +A_\tau(\pa_\la g\,\cos\Ta+\sin g\,\pa_\la\cos\Ta)\bigg]~.
\label{full}
\eea

%%%%%%%%%%%%%%%%%%%%%%%%%%%%%%%%%%%%%%%%%%%%%%%%%%%%%%%%%%%%%%%%%%%%%%
\subsection{Gauge invariance of the equations of the CW term}
%%%%%%%%%%%%%%%%%%%%%%%%%%%%%%%%%%%%%%%%%%%%%%%%%%%%%%%%%%%%%%%%%%%%%%
Unlike the quadratic and quartic Skyrme kinetic terms \re{quadSk} and \re{quartSk} which are gauge invariant as they stand,
the CW density \re{full} is not explicitly gauge invariant, so the gauge invariance of the Euler-Lagrange equations
arising from it must be checked. This is done explicitly
by subjecting \re{full} to
variations w.r.t. $A_\tau\ g,\ \Ta$ and $\F$. The results are
\bea
\de_{A_\tau}\Om_{\rm CW}&=&-2
\vep^{\tau\la\mu\nu}\bigg[(1-\cos g)\pa_\la g\,\pa_\mu\cos\Ta(A_\nu-\pa_\nu\F)\nonumber\\
&&\qquad\qquad +F_{\mu\nu}(\pa_\la g\,\cos\Ta+\sin g\,\pa_\la\cos\Ta)\bigg]
\label{dA}\\
\de_{g}\Om_{\rm CW}&=&-\vep^{\tau\la\mu\nu}\,F_{\mu\nu}
\left[(1-\cos g)\pa_\tau\cos\Ta\,(A_\la-\pa_\la\F)
+\frac12\,\cos\Ta\,F_{\tau\la}\right]
\label{dg}\\
\de_{\Ta}\Om_{\rm CW}&=&\vep^{\tau\la\mu\nu}F_{\mu\nu}
\left[(1-\cos g)\pa_\tau g\,(A_\la-\pa_\la\F)
-\frac12\,\sin g\,F_{\tau\la}\right]
\label{dTa}\\
\de_{\F}\Om_{\rm CW}&=&-\vep^{\tau\la\mu\nu}
\,(1-\cos g)\pa_\tau g\ \pa_\la\cos\Ta\ F_{\mu\nu}\,,
\label{dF}
\eea
which
are gauge invariant since the connection $A_\mu$ appears only in the combination
\[
A_\mu-\pa_\mu\F\,,
\]
with all other terms there being by definition gauge invariant.

%%%%%%%%%%%%%%%%%%%%%%%%%%%%%%%%%%%%%%%%%%%%%%%% 
%\appendix
   \setcounter{equation}{0}
   \section{The $2+2$ constraint compliant parametrisation}
\renewcommand{\theequation}{B.\arabic{equation}}
   The analysis carried out in the above Subsection is carried out here
in the formulation of Ref.~\cite{Tchrakian:2015pka}, where
instead of parametrising the Skyrme scalar by the $SU(2)$ group element $U$, the real valued four component
field $\f^a=(\f^\al,\f^A)$ of the $O(4)$ sigma model, satisfying the constraint
\[
|\f^a|^2=|\f^\al|^2+|\f^A|^2=1\ ,\quad \al=1,2\ ,\ A=3,4\,,
\]
is employed, and the definition of the covariant derivative of $\f^a$ given by \re{coval}-\re{covA}, is precisely
equivalent with the covariant derivative \re{covU} of $U$.

This is much the more natural parametrisation
of the $SO(2)$ gauged $O(4)$ sigma model scalar $\f^a$, but we have eschewed this choice and opted for the
parametrisation of Appendix {\bf A} in the main text because of technicalities associated with the numerical
construction.

The relation of the $SU(2)$ valued field $U$ to the scalar $\f^a$ is nautral and is given by
\be
\label{Uf}
U=\f^a\,\tilde\Si_a\ ,\quad{\rm and}\quad U^{\dagger}=U^{-1}=\f^a\,\Si_a\
\ee
where
\be
\label{Sisi}
\Si_\al=i\,\si_\al\ \ ,\ \Si_3=i\,\si_3\ ,\ \Si_4=\eins\quad{\rm and}\quad
\tilde\Si_\al=-i\,\si_\al\ \ ,\ \tilde\Si_3=-i\,\si_3\ ,\ \tilde\Si_4=\eins
\ee
in terms of the Pauli spin matrices $\si_\al=(\si_1,\si_2)$ and $\si_3$.

In this notation, one can conveniently express the $su(2)$ algebra valued quantities
\be
\label{UdU}
U^{-1}\pa_\mu U=-2\,\f^a\,\Si_{ab}^{(+)}\,\f^b\quad{\rm and}\quad
\pa_\mu UU^{-1}=2\,\f^a\,\Si_{ab}^{(-)}\,\f^b
\ee
where $\Si_{ab}^{(\pm)}$ are the positive and negative chiral representations of the $SO(4)$ algebra,
$\ga_{ab}=-\frac14[\ga_a,\ga_b]$,
\[
\Si_{ab}^{(\pm)}=\frac12(\eins\pm\ga_5)\,\ga_{ab}
\]
such that
\be
\label{Si+-}
\Si_{ab}^{(+)}=-\frac14\,\Si_{[a}\tilde\Si_{b]}\quad{\rm and}\quad
\Si_{ab}^{(-)}=-\frac14\,\tilde\Si_{[a}\Si_{b]}\,.
\ee

It is useful to note that $\Si_{ab}^{(\pm)}$ are self-dual and anti--self-dual respectively,
\be
\label{sdasd}
\Si_{ab}^{(+)}=\frac12\,\vep_{abcd}\ \Si_{cd}^{(+)}\quad{\rm and}\quad
\Si_{ab}^{(-)}=-\frac12\,\vep_{abcd}\ \Si_{cd}^{(-)}\,.
\ee

The natural parametrisation of $\f^\al$ and $\f^A$ is
\bea
\f^\al&=&\sin\frac{f}{2}\,n^\al\ ,\qquad n^\al=\left(\begin{array}{c}\cos\psi\\ \sin\psi\end{array}\right)\label{para1}\\
\f^A&=&\cos\frac{f}{2}\,n^A\ ,\qquad n^A=\left(\begin{array}{c}\cos\chi\\ \sin\chi\end{array}\right)\,,\label{para2}
\eea
in which according to the gauging prescription \re{coval}-\re{covA}, the functions $f$
and $n^A[\chi]$ are gauge-inert and only the unit doublet $n^\al[\psi]$ gauge-rotates.

In this parametrisation the quadratic kinetic Skyrme term is calculated immediately
to be
\be
\label{quadSk1}
|D_\mu\f^a|^2=\frac14|\pa_\mu f|^2+\cos^2\frac{f}{2}|\pa_\mu\chi|^2+\sin^2\frac{f}{2}|(A_\mu-\pa_\mu\psi)|^2
\ee 
and the quartic kinetic Skyrme term
\bea
\label{quartSk1}
|D_{[\mu}\f^aD_{\nu]}\f^b|^2&=&\cos^2\frac{f}{2}\left\{|\pa_\mu f|^2\,|\pa_\mu \chi|^2
-(\pa_\mu f\pa_\mu\chi)^2\right\}
\\
\nonumber
&&+\sin^2\frac{f}{2}\left\{|\pa_\mu f|^2\,|(A_\nu-\pa_\nu \psi)|^2
-[\pa_\mu f(A_\mu-\pa_\mu\psi)]^2\right\}
\\
\nonumber
&&+\sin^2\frac{f}{2}\left\{|\pa_\mu \chi|^2\,|(A_\nu-\pa_\nu \psi)|^2
-[\pa_\mu \chi(A_\mu-\pa_\mu\psi)]^2)\right\}
\eea
which are both gauge invariant.

Next, we proceed to calculate the CW density in this parametrisation.
Substituting \re{para1}-\re{para2} into \re{UdU}, and exploiting the (anti)--self-duality relations \re{sdasd},
one finds (after some lengthy calculation)
\bea
U^{-1}\,\pa_\mu U&=&i(X_\mu\,\si_3+Y_\mu\,\si_1+Z_\mu\,\si_2),
\label{U*dU}\\
\pa_\mu U\,U^{-1}&=&-i(\bar X_\mu\,\si_3+\bar Y_\mu\,\si_1+\bar Z_\mu\,\si_2),
\label{UdU*}
\eea
in which
\bea
X_\mu&=&\frac12\left[\pa_\mu(\psi+\chi)-\cos f\,\pa_\mu(\psi-\chi)
\right],\nonumber\\
Y_\mu&=&-\frac12\left[\pa_\mu f\,\sin(\psi+\chi)+\sin f\,\cos(\psi+\chi)\,\pa_\mu(\psi-\chi)
\right],\label{XYZ}\\
Z_\mu&=&\frac12\left[\pa_\mu f\,\cos(\psi+\chi)-\sin f\,\sin(\psi+\chi)\,\pa_\mu(\psi-\chi)
\right],\nonumber
\eea
and
\bea
\bar X_\mu&=&\frac12\left[\pa_\mu(\psi-\chi)-\cos f\,\pa_\mu(\psi+\chi)
\right],
\nonumber\\
\bar Y_\mu&=&-\frac12\left[\pa_\mu f\,\sin(\psi-\chi)+\sin f\,\cos(\psi-\chi)\,\pa_\mu(\psi+\chi)
\right],
\label{barXYZ}\\
\bar Z_\mu&=&\frac12\left[\pa_\mu f\,\cos(\psi-\chi)-\sin f\,\sin(\psi-\chi)\,\pa_\mu(\psi+\chi)
\right].
\nonumber
\eea

The result is,
\bea
\vep^{\tau\la\mu\nu}\,\mbox{Tr}Q\left[(U^{-1}\pa_\mu U)(U^{-1}\pa_\nu U)(U^{-1}\pa_\la U\right]&=&
\frac12\,\vep^{\tau\la\mu\nu}\pa_\la f\sin f\,\pa_\mu\,\psi\,\pa\chi,\nonumber\\
\vep^{\tau\la\mu\nu}\,\mbox{Tr}Q\left[(U\pa_\mu U^{-1})(U\pa_\nu U^{-1})(U\pa_\la U^{-1})\right])&=&
\frac12\,\vep^{\tau\la\mu\nu}\pa_\la f\sin f\,\pa_\mu\,\psi\,\pa\chi,\nonumber
\eea
which when substituted in \re{W1} yield the
density $\Om_{\rm CW}^{(1)}=A_\tau\,W_{(1)}^{\tau}$ appearing in \re{anomaly12}-\re{anomaly1212}
\be
\label{Om1}
\Om_{\rm CW}^{(1)}=\vep^{\tau\la\mu\nu}A_\tau\,\pa_\la f\sin f\,\pa_\mu\,\psi\,\pa_\nu\chi\,.
\ee

Next, we dispose of the term $\Om_{\rm CW}^{(2)}=W_{(2)}^{\tau\mu\nu}\,A_\tau F_{\mu\nu}$ in \re{anomaly12}-\re{anomaly1212}.
For this, one calculates $W_{(2)}^{\tau\mu\nu}$ in \re{W2}. It can be checked that the second term in \re{W2} yields
exactly the same expression as the first term, which, can be readily calculated, using
\re{U*dU}-\re{UdU*}, to give
\be
\label{Om2}
\Om_{\rm CW}^{(2)}=\frac{i^2}{2}\,\vep^{\tau\la\mu\nu}A_\tau F_{\mu\nu}\,(1+\cos f)\,\pa_\la\chi\,,
\ee
which allows the concise expression of $\Om_{\rm CW}=\Om_{\rm CW}^{(1)}+\Om_{\rm CW}^{(2)}$
\bea
\label{Om22}
\Om_{\rm CW}&=&\vep^{\tau\la\mu\nu}\left[
A_\tau\,\pa_\la f\sin f\,\pa_\mu\,\psi\,\pa_\nu\chi-\frac12\,A_\tau F_{\mu\nu}\,(1+\cos f)\,\pa_\la\chi\right]\\
&=&-\vep^{\tau\la\mu\nu}\,A_\tau\,\pa_\la\chi\left[-(\pa_\mu\cos f)\pa_\nu\psi+\frac12\,F_{\mu\nu}(1+\cos f)
\right]\,.\nonumber
\eea

To test the gauge transformation properties of the resulting Euler-Lagrange equations of \re{Om22}, one needs the
transformation properties of the functions $A_\mu,\ \psi,\ \chi$ and $f$.
It can be seen from the definition of the covariant derivatives \re{coval}-\re{covA}
and the parametrisations \re{para1}-\re{para2}, that under the $SO(2)$
gauge transformation they transform the following way
\bea
A_\tau&\to& A_\tau +\pa_\tau\La\label{a1}\\
\psi&\to&\psi+\La\label{a2}\\\
\chi&\to&\chi\label{a3}\\\
f&\to&f\,,\label{a4}\
\eea
where the functions $\chi$ and $f$ are inert under the Abelian gauge transformation.

The resulting equations of motion w.r.t. variations of $A_\tau$, $f$, $\psi$ and $\chi$, in that order are
\bea
\vep^{\tau\la\mu\nu}\pa_\la\chi\left[\frac12\,F_{\mu\nu}(1+\cos f)+(\pa_\mu\cos f)(A_\nu-\pa_\nu\psi)\right]&=&0,\label{A}
\\
\frac12\vep^{\tau\la\mu\nu}\,F_{\mu\nu}\,\pa_\la\chi\,(A_\tau-\pa_\tau\psi)&=&0,
\label{f}
\\
-\frac12\vep^{\tau\la\mu\nu}\,F_{\mu\nu}\,\pa_\la\chi\,(\pa_\tau\cos f)&=&0,
\label{psi}
\\
\vep^{\tau\la\mu\nu}\left[\frac14(1+\cos f)F_{\mu\nu}F_{\tau\la}-\frac12\,F_{\mu\nu}(\pa_\la\cos f)
\,(A_\tau-\pa_\tau\psi)\right]&=&0,
\label{chi}
\eea
each of which is gauge invariant according to \re{a1}-\re{a4}.

The analysis carried out in Section {\bf 3} employing the constraint compliant parametrisation of Appendix {\bf A}
can be carried out readily using the parametrisation given here in Appendix {\bf B}. In particular the Maxwell
equation analogous with \re{dA1} there and the corresponding $\F$ equation with
\re{dF1}, follow from \re{A} and \re{psi} respectively for the variations of $A_\mu$ and $\psi$. The role played
there by $\F$ in the definition of the Noether (electric) charge is played by the function $\psi$.

As for imposition of axial symmetry, here in analogy with \re{SkSc} we have 
\be
\label{SkScx}
f(x_\mu)=f(\rho,z)\ ,\quad\chi(x_\mu)=\chi(\rho,z)\ ,\quad\psi(x_\mu)=n\,\vf
\ee
where $\vf$ is the azimuthal angle in the $(x_1,x_2)$ plane and $n$ is an integer.

The rest of the analysis in Section {\bf 4} follows systematically with qualitatatively similar results. The 
numerical work in Section {\bf 4} however is carried out using exclusively the $3+1$ parametrisation of
Appendix {\bf A}. This is because the numerical construction even of the (ungauged) $1$-Skyrmion in the
$2+2$ parametrisation here is far from straightforward.

%%%%%%%%%%%%%%%%%%%%%%%%%%%%%%%%%%%%%%%%%%%%%%%%%%%%%%%%%%%%%%%%%%
\begin{small}

\end{small}

\end{document}